\documentclass[draft]{agujournal2019}
\usepackage{url} %this package should fix any errors with URLs in refs.
\usepackage{lineno}
\usepackage[inline]{trackchanges} %for better track changes. finalnew option will compile document with changes incorporated.
\usepackage{soul}
\journalname{Space Weather}

\begin{document}

%%%%%%%%%%%%%%%%%%%%%%%%%%%%%%%%%%%%%%%%%%%%%%%
%  TITLE
%
% (A title should be specific, informative, and brief. Use
% abbreviations only if they are defined in the abstract. Titles that
% start with general keywords then specific terms are optimized in
% searches)
%
%%%%%%%%%%%%%%%%%%%%%%%%%%%%%%%%%%%%%%%%%%%%%%%

% Example: \title{This is a test title}

\title{Solar Wind with Field Lines and Energetic Particles (SOFIE) Model: Application to Historical Solar Energetic Particle Events}

%%%%%%%%%%%%%%%%%%%%%%%%%%%%%%%%%%%%%%%%%%%%%%%
%
%  AUTHORS AND AFFILIATIONS
%
%%%%%%%%%%%%%%%%%%%%%%%%%%%%%%%%%%%%%%%%%%%%%%%

% Authors are individuals who have significantly contributed to the
% research and preparation of the article. Group authors are allowed, if
% each author in the group is separately identified in an appendix.)

% List authors by first name or initial followed by last name and
% separated by commas. Use \affil{} to number affiliations, and
% \thanks{} for author notes.
% Additional author notes should be indicated with \thanks{} (for
% example, for current addresses).

% Example: \authors{A. B. Author\affil{1}\thanks{Current address, Antartica}, B. C. Author\affil{2,3}, and D. E.
% Author\affil{3,4}\thanks{Also funded by Monsanto.}}

\authors{Lulu Zhao\affil{1}, Igor Sokolov\affil{1}, Tamas Gombosi\affil{1}, David Lario\affil{2}, Kathryn Whitman\affil{3,4}, Zhenguang Huang\affil{1}, Gabor Toth\affil{1}, Ward Manchester\affil{1}, Bart van der Holst\affil{1},  Nishtha Sachdeva\affil{1}}

\affiliation{1}{Department of Climate and Space Sciences and Engineering, University of Michigan, Ann Arbor, MI, 48103, USA}
\affiliation{2}{NASA Goddard Space Flight Center, 8800 Greenbelt Rd, Greenbelt, MD 20771, USA}
\affiliation{3}{University of Houston, 4800 Calhoun Rd, Houston, TX, 77204, USA}
\affiliation{4}{KBR, 601 Jefferson Street, Houston, TX, 77002, USA}

% Corresponding author mailing address and e-mail address:

% (include name and email addresses of the corresponding author.  More
% than one corresponding author is allowed in this LaTeX file and for
% publication; but only one corresponding author is allowed in our
% editorial system.)

% Example: \correspondingauthor{First and Last Name}{email@address.edu}

\correspondingauthor{Lulu Zhao}{zhlulu@umich.edu}

%%%%%%%%%%%%%%%%%%%%%%%%%%%%%%%%%%%%%%%%%%%%%%%
% KEY POINTS
%%%%%%%%%%%%%%%%%%%%%%%%%%%%%%%%%%%%%%%%%%%%%%%
%  List up to three key points (at least one is required)
%  Key Points summarize the main points and conclusions of the article
%  Each must be 140 characters or fewer with no special characters or punctuation and must be complete sentences

% Example:
% \begin{keypoints}
% \item	List up to three key points (at least one is required)
% \item	Key Points summarize the main points and conclusions of the article
% \item	Each must be 140 characters or fewer with no special characters or punctuation and must be complete sentences
% \end{keypoints}

\begin{keypoints}
\item Solar Energetic Particles
\item Space Radiation Prediction
\item Space Weather Forecast
\end{keypoints}

%%%%%%%%%%%%%%%%%%%%%%%%%%%%%%%%%%%%%%%%%%%%%%%
%
%  ABSTRACT and PLAIN LANGUAGE SUMMARY
%
% A good Abstract will begin with a short description of the problem
% being addressed, briefly describe the new data or analyses, then
% briefly states the main conclusion(s) and how they are supported and
% uncertainties.

% The Plain Language Summary should be written for a broad audience,
% including journalists and the science-interested public, that will not have 
% a background in your field.
%
% A Plain Language Summary is required in GRL, JGR: Planets, JGR: Biogeosciences,
% JGR: Oceans, G-Cubed, Reviews of Geophysics, and JAMES.
% see http://sharingscience.agu.org/creating-plain-language-summary/)
%
%%%%%%%%%%%%%%%%%%%%%%%%%%%%%%%%%%%%%%%%%%%%%%%

\begin{abstract}
In this paper, we demonstrate the applicability of the data-driven and self-consistent solar energetic particle model, Solar-wind with FIeld-lines and Energetic-particles (SOFIE), to simulate acceleration and transport processes of solar energetic particles.
SOFIE model is built upon the Space Weather Modeling Framework (SWMF) developed at the University of Michigan. In SOFIE, the background solar wind plasma in the solar corona and interplanetary space is calculated by the Aflv\'en Wave Solar-atmosphere Model(-Realtime) (AWSoM-R)  driven by the near-real-time hourly updated Global Oscillation Network Group (GONG) solar magnetograms. In the background solar wind, coronal mass ejections (CMEs) are launched by placing an imbalanced magnetic flux rope on top of the parent active region, using the Eruptive Event Generator using Gibson-Low model (EEGGL). The acceleration and transport processes are modeled by the Multiple-Field-Line Advection Model for Particle Acceleration (M-FLAMPA). 
In this work, nine solar energetic particle events (Solar Heliospheric and INterplanetary Environment (SHINE) challenge/campaign events) are modeled. 
The three modules in SOFIE are validated and evaluated by comparing with observations, including the steady-state background solar wind properties, the white-light image of the CME, and the flux of solar energetic protons, at energies of $\ge$ 10 MeV. 

\end{abstract}

\section*{Plain Language Summary}
In this paper, we describe one physics-based solar energetic particle model, called Solar-wind with FIeld-lines and Energetic-particles (SOFIE). This model is designed to simulate the acceleration and transport processes of solar energetic particles in the solar atmosphere and interplanetary space. 
SOFIE is built on the Space Weather Modeling Framework (SWMF) developed at the University of Michigan. 
There are three modules in the SOFIE model, the background solar wind module, the coronal mass ejection (CME) initiation and propagation module, and the particle acceleration and transport module.
The background solar wind plasma in the solar corona and interplanetary space is modeled by the Aflv\'en Wave Solar-atmosphere Model(-Realtime) (AWSoM-R)  driven by the near-real-time hourly updated Global Oscillation Network Group (GONG) solar magnetograms. In the background solar wind, the CMEs are launched by placing an unbalanced magnetic flux rope on top of the active region, using the Eruptive Event Generator using Gibson-Low configuration (EEGGL). The acceleration and transport processes are then modeled self-consistently by the Multiple-Field-Line Advection Model for Particle Acceleration (M-FLAMPA). 
Using SOFIE, we modeled nine historical solar energetic particle events. The performance of the model and its capability in making space radiation prediction is discussed.

\section{Introduction}
Solar energetic particles (SEPs) can be accelerated over a wide range of energies extending up to GeVs. 
%At relatively low energies (e.g., 10 MeV), their flux intensity can exceed the background of galactic cosmic rays by several orders of magnitude. 
They are hazardous not only to humans in space but also to electronics and other sensitive components of spacecraft affecting their operations.
Protons of $>\!\!100$ MeV with elevated fluxes exceeding 1 proton flux unit (pfu) are responsible for an increased astronaut exposure inside spacecraft shielding, and protons of $>\!\!150$ MeV are very difficult to shield against as they can penetrate 20 gm cm$^{-2}$ (7.4 cm of Al, or 15.5 cm of water/human tissue) \cite<e.g.>{Reames2013}. 
Furthermore, $>500$ MeV protons can penetrate the atmosphere and pose radiation hazards to aviation. Besides protons, energetic heavy ions can also be of severe radiation concerns.
Therefore, a reliable prediction of the timing and absolute flux of energetic protons above different energies is needed to provide support for future space exploration. 
However, the sparsity and large variability of SEP events make them difficult to predict. 

Many currently-existing SEP prediction models use post-eruptive observations of solar flares/CMEs to predict SEP events \cite<e.g.>{Balch2008,Smart1976,Smart1989,Smart1992,Inceoglu2018,Huang2012,Belov2009,Garcia2004,Laurenza2009,Richardson2018}.
There are also models that make predictions of the eruptive events (flares, CMEs, SEPs) using solar magnetic field measurements \cite{Georgoulis2008, Park2018, Bobra2016, Bobra2015, Huang2018, Boucheron2015, Falconer2014, Bloomfield2012, Colak2009,Papaioannou2015, Anastasiadis2017,Engell2017,Garcia-Rigo2016,Tiwari2015,Kasapis2022}. 
In addition, because of the shorter transit times of relativistic electrons or very high energy protons compared to $\sim\!\!10$ MeV protons, near-real-time observations of $\sim$MeV electrons \cite{Posner2007} and/or $>\!\!100$MeV protons \cite{Boubrahimi2017, Nunez2015, Nunez2011} have also been used to predict the arrival of $>\!\!10$ MeV protons. 

A recent review by \citeA{Whitman2022} summarizes more than three dozen SEP models to predict the occurrence probability and/or properties of SEP events. 
In \citeA{Whitman2022}, three approaches of the prediction models are discussed, empirical, machine learning (ML) and physics-based models. 
Empirical and ML models are built upon potential causality relations between the observable and predictable and they can make rapid predictions, often within seconds or minutes after the input data becoming available. Such models hold value as they can generally issue forecasts prior to the peak of an SEP event. However, since empirical and ML models are built upon historic events, it is difficult to validate their predictions at locations where no routine/historical observations have been made, e.g., the journey from Earth to Mars. 
And predictions can only be made for the specific energy channels upon which these models are built/trained. These models may also have difficulty in predicting extreme events since there are few such events available for training \cite<e.g.>{Bain2021, Nunez2015, Whitman2022}.
On the other hand, physics-based models are based on first principles \cite{Tenishev2021, schwadron2010, Alberti2017, Alho:2019, Marsh2015, Hu2017, Sokolov2004, Borovikov2018, Wijsen2020, Wijsen2022, Li2021, Luhmann2007, Aran2017, Strauss2015, Kozarev2017, Kozarev2022, Linker2019, Zhang2017}.
Physics-based models are usually computationally expensive, and in order for the physics-based models to make meaningful predictions, they need to run faster than real-time. Moreover, many of the underlying physical mechanisms involved in the development of SEP events are still under-debate, including the particle acceleration processes in the low corona, the particle's interaction with turbulence magnetic field in the heliosphere, and the seed particles that are injected into the particle acceleration processes.
However, physics-based models are still highly attractive, since they solve the acceleration and transport processes of energetic particles and therefore they are able to provide time profiles and energy spectra of SEPs at any location of interest in the heliosphere. 

In this work, we demonstrate our attempt to model and make potential predictions of the energetic protons by using the self-consistent physics-based model, called SOlar wind with FIeld lines and Energetic particles (SOFIE). 
In this paper, we will apply the SOFIE model to nine historical SEP events. These nine SEP events are chosen from the Solar Heliospheric and INterplanetary Environment (SHINE) challenge/campaign events, which were selected based on their elevated intensities that were relevant to operations\footnote{\url{https://ccmc.gsfc.nasa.gov/challenges/sep/shine2018/}, \url{https://ccmc.gsfc.nasa.gov/challenges/sep/shine2019/},
\url{https://ccmc.gsfc.nasa.gov/community-workshops/ccmc-sepval-2023/}}.

\section{SOFIE}
In order to build a physics-based SEP model, a background solar wind module, a CME generation and propagation module, and
a particle acceleration and transport module are required. 
In SOFIE, the background solar wind plasma in the solar corona and interplanetary space is modeled by the Alfv\'en Wave Solar-atmosphere Model(-Realtime) (AWSoM-R)  driven by hourly solar magnetograms obtained from the Global Oscillation Network Group (GONG) of the National Solar Observatory (NSO).
CMEs are launched by placing an imbalanced magnetic flux rope on top of the parent active region, using the Eruptive Event Generator using Gibson-Low configuration (EEGGL). 
The acceleration and transport processes of energetic particles are then modeled by the Multiple-Field-Line-Advection Model for Particle Acceleration (M-FLAMPA). 
All the three modules are fully integrated through the Space Weather Modeling Framework (SWMF) developed at the University of Michigan.
In this section, we briefly introduce each module. 

\subsection{Background Solar Wind}
The 3D global solar wind plasma in the Solar Corona (1 $R_s$ - 20 $R_s$) and inner heliosphere (20 $R_s$ - 5 AU) is modeled by using  AWSoM-R as configured in the SWMF \cite{Sokolov2013, Sokolov2021, Gombosi2018, Gombosi2021}.
AWSoM-R is an Aflv\'en wave-driven, self-consistent solar atmosphere model, in which the coronal plasma is heated by the dissipation of two discrete turbulence populations propagating parallel and antiparallel to the magnetic field \cite{Sokolov2013}.
The AWSoM-R solar wind model has been validated by comparing simulations and observations of both the in-situ macroscopic properties of the solar wind and the line-of-sight (LoS) appearance of the corona as observed in different wavelengths \cite{Sachdeva2019, Gombosi2021}.
The inner boundary of AWSoM-R is characterized by the magnetic field measurement made by either ground-based or space-based observatories. 
In all the SEP events we modeled in this work, hourly-updated GONG solar magnetograms are used.\footnote{\url{https://gong.nso.edu/data/magmap/}}

A validated background solar wind solution is critical in modeling the transport processes of energetic particles as it provides the magnetic field configuration where particles propagate, allowing the computation of the energetic particle properties observed by spacecraft at specific heliospheric locations.
Numerical solutions of the full set of ideal or resistive magnetohydrodynamic (MHD) equations so far have not been able to reproduce aligned interplanetary stream lines and magnetic field lines in corotating frames. One of the reasons for this discrepancy is the numerical reconnection across the heliospheric current sheet: the reconnected field is directed across the current sheet, while the global solar wind streams along the current sheet, thus resulting in ``V-shaped'' magnetic field lines and significant misalignment between field lines and stream lines.
It is impossible to follow particles' trajectory in ``V-shaped'' magnetic field lines, therefore, stream lines are usually used instead \cite{Young2020}.
Within regular MHD, there is no mechanism to re-establish the streamline-fieldline alignment.
Recently, \citeA{Sokolov2022} introduced the Stream-Aligned MHD method that ``nudges'' the magnetic field lines and plasma stream lines towards each other. A detailed explanation and illustration of this method is discussed in \citeA{Sokolov2022}.
In SOFIE, we will solve Stream-Aligned MHD to get a steady state solar wind plasma background representative of the pre-event ambient solar wind and magnetic medium where CMEs and SEPs propagate.

\subsection{CME Initiation and Propagation}
The CME generation in SOFIE is modeled by the EEGGL module in SWMF \cite{Manchester2004,Manchester2004a, Manchester2006, Manchester2014a, Manchester2014b, Lugaz2005b, Lugaz2007, Kataoka2009, Jin2016, Jin2017b,Shiota2016,Borovikov2017}.
The initial conditions of the CME within the solar corona is treated by inserting an unstable (or force imbalanced) flux rope suggested by \citeA{Gibson1998} into an active region.
The magnetogram from GONG and the observed CME speed (from Coordinated Data Analysis Web (CDAW) catalog and/or The Space Weather Database Of Notifications, Knowledge, Information (DONKI) database) are used to calculate the flux rope parameters.
This approach offers a relatively simple, and inexpensive model for CME initiation based on empirical features of pre-event conditions \cite<e.g.>{Gombosi2021}.
The EEGGL module is publicly available for download at \url{http://csem.engin.umich.edu} or can also be used through the website of the Community Coordinated Modeling Center (CCMC, \url{https://ccmc.gsfc.nasa.gov/eeggl/}).
The subsequent propagation of CMEs in the solar corona and interplanetary medium are modeled using the AWSoM-R module. 
The EEGGL model to initialize CMEs and the subsequent CME/ICME evolution has been extensively used and validated \cite<e.g.>{Jin2017, Manchester2017b, Manchester2014a, Manchester2004a, Manchester2012a, Manchester2005a, Manchester2008a, Manchester2014a, Roussev2004a, Roussev2008a, vanderHolst2009a, vanderHolst2007a}. 

\subsection{Particle Tracker}
In SOFIE, protons are accelerated at the shocks driven by CMEs through first order Fermi acceleration mechanism \cite{Krymsky1977, Axford1977, blandford78, Bell1978a, Bell1978b}. 
The acceleration and transport processes are modeled by the M-FLAMPA module in SWMF.
In M-FLAMPA, the time-evolving magnetic field lines are extracted from the AWSoM-R solutions, along which the particle distribution functions are solved, following the Parker diffusion equation \cite{Sokolov2004,Borovikov2018}.
Novel mathematical methods are applied to the extracted magnetic field lines to sharpen the shocks thus making the Fermi acceleration process to be more efficient \cite{Sokolov2004}.
The injection of suprathermal protons into the CME-driven shock acceleration system is described in \citeA{Sokolov2004}. 
The interaction between the energetic protons and turbulent magnetic fields is modeled by the diffusion processes along the background magnetic field lines. 
The diffusion coefficient close to the shock region is calculated self-consistently through the total Aflv\'en wave intensities obtained in the MHD simulation, and a Kolmogorov spectrum with an index of $-5/3$ is assumed. The diffusion coefficient upstream of the shock is calculated by assuming a constant mean free path. Detailed parameter settings will be discussed in Section~\ref{sec:result}.

\section{Overview of the Nine SEP Events}\label{sec:overview}
The nine SHINE challenge events were primarily chosen because they were large SEP events that were relevant to operations. 
Specifically, the 2012 July 12 event was selected because there was a large particle enhancement at Mars.
In this section, we describe the observational facts of the nine SEP events.
Table~\ref{tbl:overview} summarizes the observational facts of the CMEs and solar flares associated with the solar origin of the nine events. From left to right, each column shows the SEP event date used to identify the event, the associated CME onset time, the CME speed, the soft X-ray flare class and onset time, the NOAA active region locations on the Sun, and the NOAA active region (AR) number. The CME onset time is estimated from observations made by the Large Angle and Spectrometric Coronagraph (LASCO) instrument on board Solar \& Heliospheric Observatory (SOHO). 
Note that all the CMEs associated with the SEP events modeled in this work are categorized as halo CME in the SOHO LASCO  CME catalog CDAW\footnote{\url{https://cdaw.gsfc.nasa.gov/CME_list/halo/halo.html}}.
Each individual SEP event has been studied extensively by many papers as described below.
Key features of each individual event are as follows:

 \begin{table}
 \caption{Observational facts of the nine SEP events}
 \label{tbl:overview}
 \centering
 \begin{tabular}{ccccc}
 \hline
  Event Date  & CME Onset Time$^{a}$ & CME Speed$^{b}$ & SXR GOES Flare & NOAA AR\\
   &  [UT] & [km/s] & Class/Onset [UT] &  \\
 \hline
   2012-Mar-07 & 2012-Mar-07 00:24 & 2040 & X5.4/00:02 & N17E15(11429) \\
   2012-May-17 & 2012-May-17 01:37 & 1263 & M5.1/01:25 & N12W89(11476) \\
   2012-Jul-12 & 2012-Jul-12 16:54 & 1400 & X1.4/15:37 & S14W02(11520) \\
   2013-Apr-11 & 2013-Apr-11 07:24 & 743  & M6.5/06:55 & N09E12(11719) \\
   2014-Jan-07 & 2014-Jan-07 18:12 & 2048 & X1.2/18:04 & S15W11(11943) \\
   2017-Jul-14 & 2017-Jul-14 01:25 & 750  & M2.4/01:07 & S09W33(12665) \\
   2017-Sep-04 & 2017-Sep-04 20:24 & 1323 & M5.5/20:12$^{c}$ & S08W16(12673)\\
   2017-Sep-06 & 2017-Sep-06 12:12 & 1816 & X9.3/11:53 & S08W34(12673)\\
   2017-Sep-10 & 2017-Sep-10 15:48 & 2087 & X8.2/15:35 & S08W88(12673)\\
 \hline
\multicolumn{5}{l}{$^{a}$ The onset time is obtained from the SHINE challenge websites and visually examined.} \\
\multicolumn{5}{l}{to match the SOHO observations.} \\
\multicolumn{5}{l}{$^{b}$The CME speed is provided by the SHINE challenge website.} \\
\multicolumn{5}{l}{$^{c}$Based on inspection of SDO/AIA images.}
 \end{tabular}
 \end{table}
 
\textbf{2012-Mar-07 Event}: 
The solar origin of this SEP event is temporally associated with a X5.4 class X-ray from the NOAA Active Region (AR) 11429 at N17E15. At 00:24 UT, a fast halo CME with a plane-of-sky speed of 2040 km s$^{-1}$ was detected in LASCO/C2 coronagraph images. 
At 01:05 UT, a second flare with a class of X1.3 erupted from the same active region and a slower halo CME with a speed of 1825 km s$^{-1}$ was detected.
Detailed analyses of these two eruptions can be found elsewhere \cite<e.g.>{Patsourakos2016}.
The fact that the first CME was faster than the second CME and that the electron intensities measured by the MErcury Surface,
Space ENvironment, GEochemistry, and Ranging (MESSENGER) at 0.31 AU peaked before the occurrence of the second flare \cite<c.f. Figure~6 in>{Lario2013} suggest that the main contributor to the observed SEP event was the first solar eruption. In fact, in the analysis of SEP events observed by the two spacecraft of the Solar Terrestrial Relations Observatory (i.e., Solar TErrestrial RElations Observatory (STEREO)-Ahead and STEREO-Behind) and near-Earth spacecraft, \citeA{Richardson2014} and \citeA{Kouloumvakos2016} concluded that the first flare/CME was responsible for the SEP event at all three locations.
Therefore, in the simulation, we will consider only the first CME.
Yet the energetic particle measurement made by Geostationary Operational Environmental Satellite (GOES) shows two clear onset phases, which may correspond to the two CMEs. The peak and decay phases of the intensity profile was indistinguishable. 
%In this work, we will focus on the particle acceleration in the first CME.  
%The second population of energetic particles from the second eruption will be evaluated when analyzing the simulation results.

\textbf{2012-May-17 Event}: 
This event was the first Ground Level Enhancement (GLE) of solar cycle 24 with 
$>$433 MeV proton intensity enhancements detected by GOES-13 and up to $^{>}_{\sim}$7 GeV as inferred from neutron monitor observations \cite{Balabin2013, Li_Chuan_2013}.
This GLE, designated as GLE71, had the peculiarity of having a highly anisotropic onset as detected by several neutron monitor stations \cite{Mishev2014}.
By assuming that relativistic protons propagated scatter-free along nominal interplanetary field lines, \citeA{Li_Chuan_2013} estimated that $\sim$1.12 GeV protons were release at about 01:39$\pm$00:02 UT, in accordance with a type II radio burst and prominence eruption at the origin of the associated fast CME, and corresponding to a a height of the CME at $\sim$3.07 R$_{s}$.
It is worth noting that \citeA{Shen2013} reported two CME eruptions from the same active region that were separated by about 2 minutes. However, in the time intensity profiles of energetic protons detected by GOES, the two eruptions were not well separated. In this work, we will only consider the first CME eruption as the main accelerators of energetic particles. The same approach was adopted by \citeA{Li2021} who modeled this event using AWSoM and improved Particle Acceleration and Transport in the Heliosphere model (iPATH) models.

\textbf{2012-Jul-12 Event}: 
The CME at the origin of this SEP event generated the fourth strongest geomagnetic storm of solar cycle 24 \cite{Gil2020}.
The prompt component of this SEP event showed $>$100 MeV proton intensity enhancements as observed by GOES-13 \cite<c.f. Figure~6 in>{Gil2020} and the arrival of the shock at 1 AU driven by the CME was accompanied by a strong energetic storm particle (ESP) event \cite<e.g.>{Wijsen2022}.
Details of the solar eruption that generated this event, reconstructions of the CME structure as observed by coronagraphs, and the topology of the CME at its arrival at 1 AU can be found in \citeA{Scolini2019}, \citeA{Gil2020} and references therein.
%At 15:37 UT 2012 Jul 12, a X1.4 class flare erupts from active region 11520 located at S15W02. The CME followed the flare occurred at 16:54 UT 2012 Jul 12 with a speed of 1400 km s$^{-1}$. 

\textbf{2013-Apr-11 Event}: 
This SEP event was the first Fe-rich event
of solar cycle 24 as evidenced by ion data collected by STEREO-B and near-Earth spacecraft \cite{Cohen2014}. 
The filament eruption origin of the CME that generated this SEP event has been studied by several authors \cite<e.g.>{Vermareddy2015, Joshi2017, Fulara2019}.
The EUV wave associated with the origin of this event propagated
mostly toward the footpoint of the nominal interplanetary magnetic field line connecting to STEREO-B,
but signatures of the EUV wave reaching the
footpoints of the interplanetary magnetic field lines connecting to either STEREO-A or near-Earth spacecraft were not observed \cite{Lario2014}.
The non-arrival of the EUV wave at the magnetic footpoint of a given spacecraft does not preclude the observation of SEPs by such a spacecraft. 
\citeA{Lario2013} concluded that observation of particles by near-Earth spacecraft was due to the CME-driven shock expanding at higher altitudes
over a wide range of longitudes, without leaving an observable EUV trace in the low corona, being able to accelerate and inject particles onto the field lines connecting to near-Earth locations. 

%At 2013 Apr 11 06:55 UT, a M6.5 class flare erupts the active region 11719 located at N09E12 and at 2013 Apr 11 07:24 UT, a CME with a speed of 743 km s$^{-1}$ appears in the SOHO/LASCO C2 detector. The period before this event is very quiet.
%There is no $>$ 600 km s$^{-1}$ CMEs observed for a period of 10 days before the onset of this SEP event and no X-ray flares of class above M were observed for a period of 6 days prior to the event \cite{Lario2014}.
%The energetic proton intensity measurement shows a rapid onset followed by a gradual decay in both STEREOB and L1 locations.

%The SEP event on 2014-01-06 was clamed to be a new GLE \cite{Thakur2014} with $>$700 MeV proton intensity enahcenments observed by GOES, but \citeA{Kuhl2015} determined that SOHO did not observe a significant increase of $>$600 MeV proton intensity.

\textbf{2014-Jan-07 Event}: 
The solar eruption at the origin of the CME associated with the SEP event was analyzed in detail by \citeA{Mostl2015}.
They showed that the CME was “channeled” by strong nearby active region magnetic fields and open coronal fields into a non-radial propagation direction within $\sim$2.1 R$_{S}$, in contrast to deflection in interplanetary space. 
This phenomenon will be discussed in more detailed in Section~\ref{sec:result}, where a white-light coronagraph comparison between the simulation and observation is discussed.
\citeA{Mays2015} studied the propagation of this CME up to 1 AU and determined that only a glancing CME arrival was observed at Earth. 
The SEP intensity enhancement occurred on the tail of a very energetic SEP event with onset on 2014 January 6 \cite<see details in, e.g.,>{Thakur2014, Kuhl2015}.

%{\it old text:} At 18:04 UT 2014 Jan 7, a X1.2 class flares was detected. However, this flare was not associated with currently named NOAA active regions according to \url{https://www.solarmonitor.org/full_disk.php?date=20140107&type=saia_00193&region=}. We carefully examined the SDO/AIA movies during this period and found the brightening of the coronal loops connecting the active region 11944 and 11943 is the corresponding source. In this work, we assigned the active region to be 11943 and placed the flux rope in the corresponding flaring location. \lulu{How are the 11944 and 11943 boundaries defined?}
%The subsequent CME occurred at 18:12 UT 2014 Jan 7 with a speed of 2048 km s$^{-1}$.

\textbf{2017-Jul-14 Event}: The origin of this event was associated with a 
medium-sized (M2.4) long-duration (almost two hours) flare from a large active region that displayed a sigmoidal configuration associated with a filament/flux rope. 
A high-lying coronal EUV loop was seen moving outward, which was immediately followed by the impulsive phase of the flare \cite{Jing2021}.
The formation of the sigmoidal filament/flux rope, its expansion, and the evolution of the photospheric magnetic field, leading to the eruption of the filament and the resulting CME have been studied in detail by \citeA{James2020} (see their Figure~13). 
The arrival of the shock at Earth, accompanied by local particle intensity increases at energies $^{<}_{\sim}10$ MeV, generated a geomagnetic storm K$_{p}$=6.

%At 01:07 UT 2017 Jul 14, a M2.4 class flare was observed in active region 12665 located at S07W33. The CME later appears in the C2 field of view at 01:25 UT 2017 Jul 14 with a speed of 750 km $s^{-1}$.

\textbf{2017-Sep-04 Event}: This SEP event, together with the following two SEP events, are a series of SEP events that occurred in early September 2017, towards the end of solar cycle 24.
The solar eruptions associated with the origin of these events and their geomagnetic effects were analyzed by \citeA{Chertok2018} and \citeA{Shen2018} and references therein, whereas the resulting SEP events were described by \citeA{Bruno2019} among others.
The flare associated with the first SEP event occurred at 20:12~UT on 2017 Sep 4 and the CME occurred at 20:24~UT with a speed of 1323 km s$^{-1}$. The active region (AR 12673) was located at S09W16. The flare onsets time was estimated from the SDO/AIA movies. 
%The flare information is missing from the solar monitor website. 
From SOHO/LASCO C2 images, around two hours before the eruption of the CME associated with the SEP event, there was a preceding CME at 18:48 UT on 2017 Sep 4 with a speed of 597 km s$^{-1}$ (CDAW). From the point of view of SOHO/LASCO, the first CME propagates to the west whereas the second faster CME propagates toward the southwest. The second CME overtook the previous CME shortly after its eruption, around 21:24 UT. In this work, we attribute the main acceleration of protons to the second CME, which is faster and stronger.

\textbf{2017-Sep-06 Event}: 
A X9.3 class flare occurred at 11:54 UT on 2017 Sep 6 from the same active region AR 12673 as the 2017-Sep-04 event. At this time, the active region was located at S08W34. The CME has a speed of 1816 km s$^{-1}$. The occurrence of this SEP event was in the decay phase of the previous event, making the identification of the onset of the energetic proton intensity enhancements at different energies difficult.
%resulting a missing onset phase of the energetic protons.

\textbf{2017-Sep-10 Event}: 
At 15:35 UT on 2017 Sep 10, the same active region AR 12673 produced a X8.2 class flare. The active region rotated to S08W88. The corresponding CME has a speed of 2087 km s$^{-1}$. This event is an GLE event, GLE \#72. This event was also well-studied by multiple groups \cite<see details in>{Ding2020, Zhu2021}.

\section{SOFIE Results}\label{sec:result}

In this section, we present the results of the SOFIE model in simulating the nine SEP events. When modeling each event, we first run the AWSoM-R model to get a steady state solution of the background solar wind. In doing so, the hourly GONG magnetogram measured right before the flare eruption is chosen to drive the AWSoM-R model. The simulation domain extends from $1.105$ solar radius (Rs) to $2.5$ AU. In Section~\ref{sec:background}, we discuss the background solar wind solutions for each event and compare them with in-situ observations made by near-Earth instruments. 
After getting the steady state solar wind solution, an imbalanced magnetic flux rope is placed on top of the active region where the CME erupted from. In Section~\ref{sec:cme}, we show the 3D topology of the magnetic flux rope and compare the white-light coronagraph images calculated from simulation with the LASCO/C2 observations.
In Section~\ref{sec:mflampa}, we show the 2D spatial distribution of energetic particles in a sphere around Earth and the extracted proton flux time profiles.

\subsection{Background Solar Wind}\label{sec:background}
\begin{figure}
\noindent\includegraphics[width=0.33\textwidth]{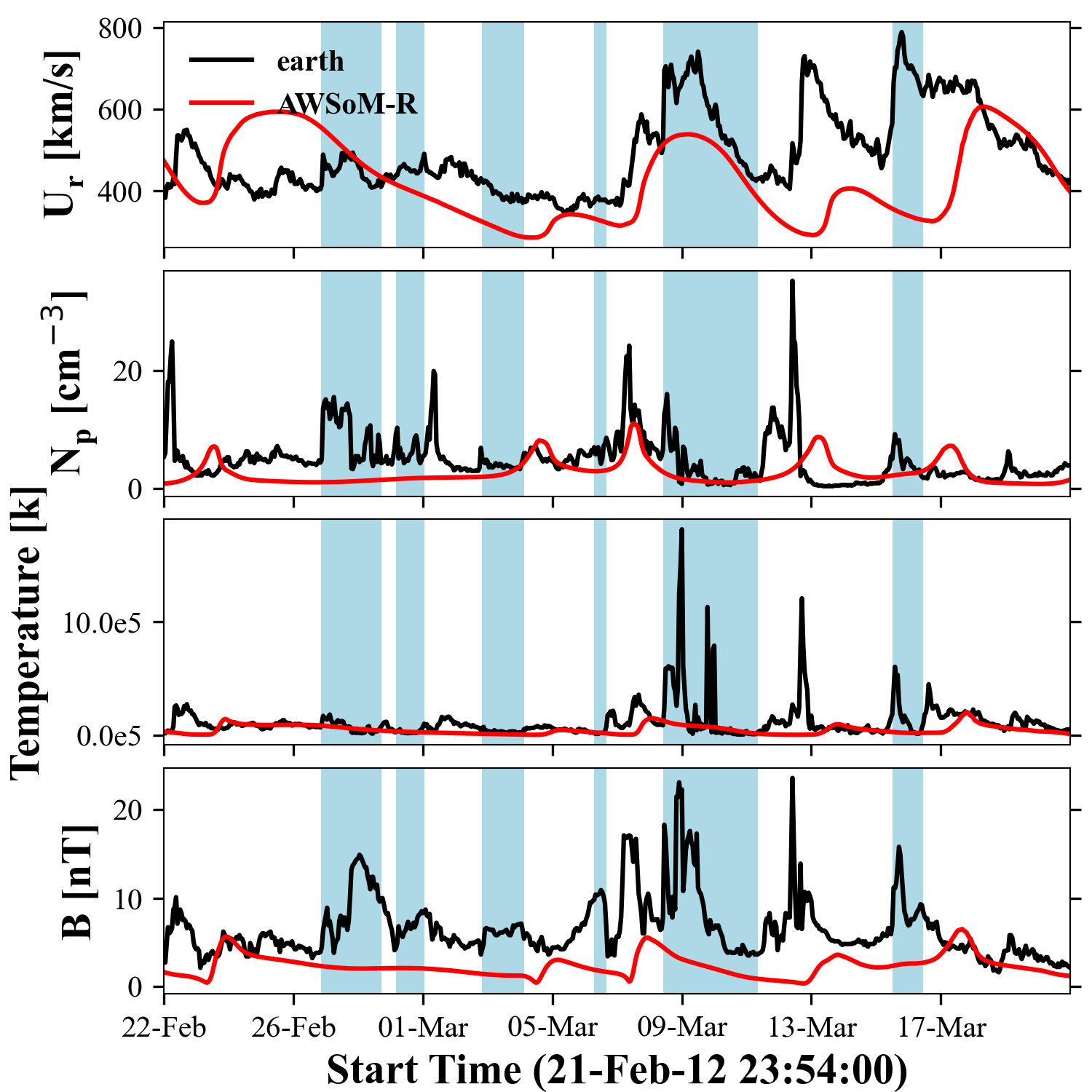}
\noindent\includegraphics[width=0.33\textwidth]{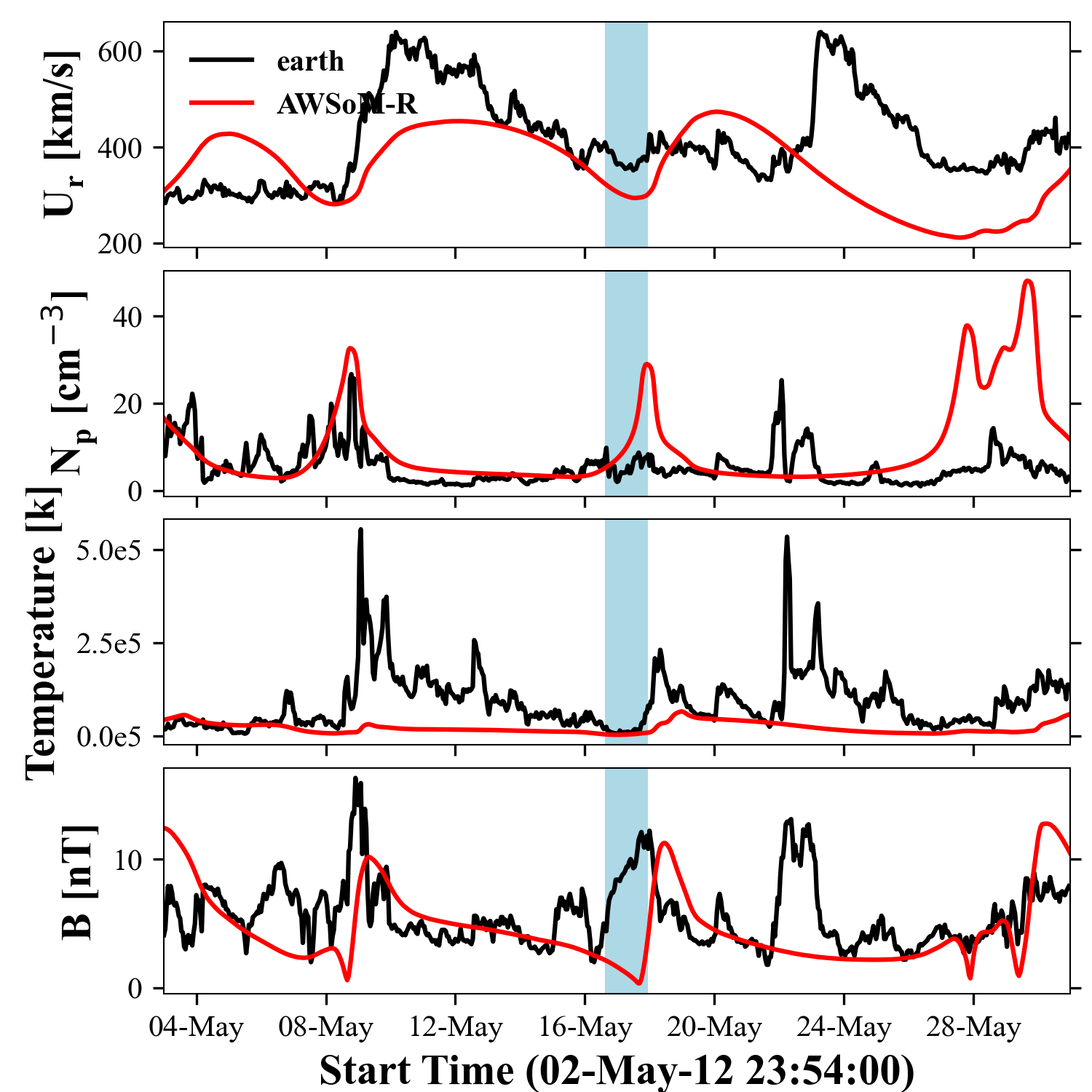}
\noindent\includegraphics[width=0.33\textwidth]{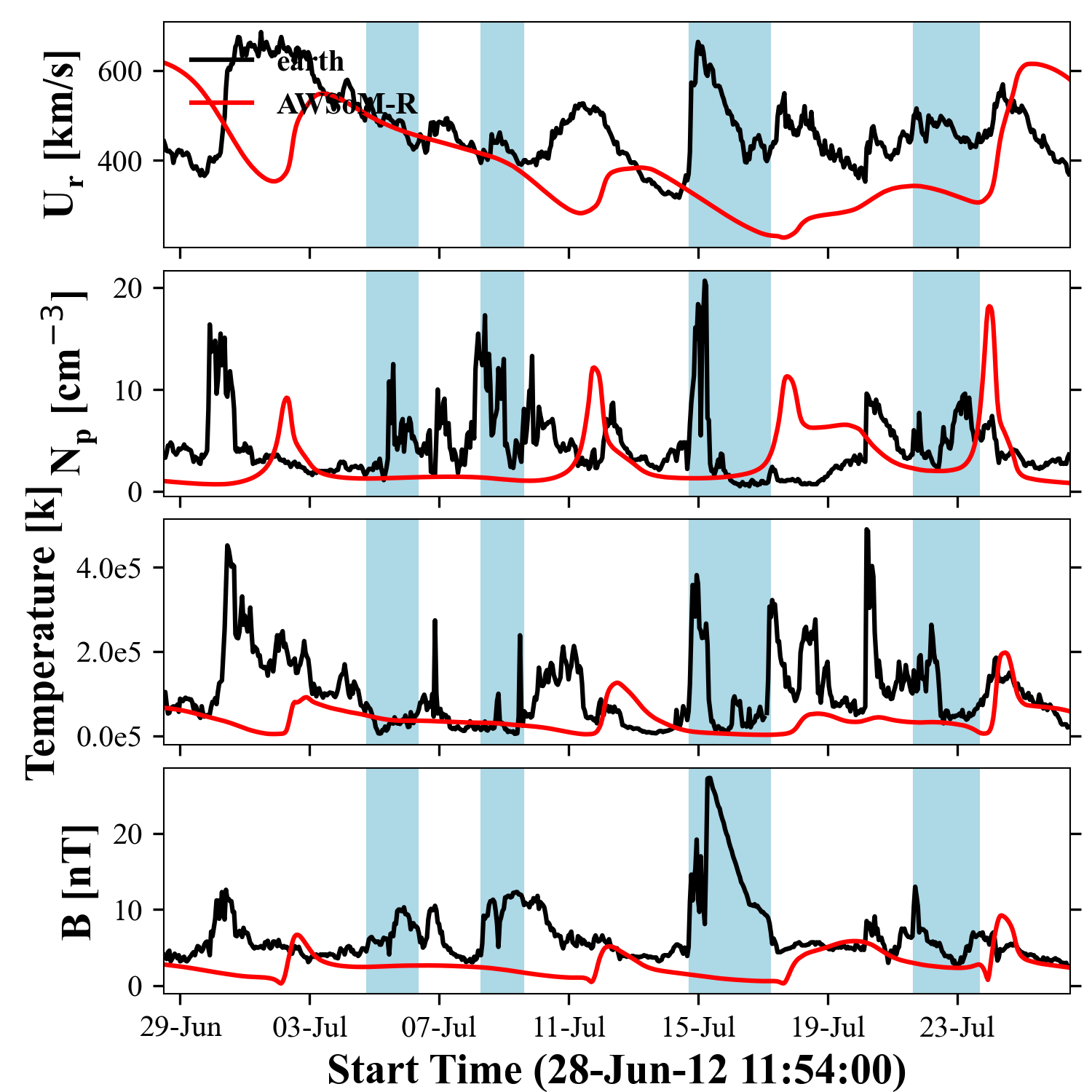}
\noindent\includegraphics[width=0.33\textwidth]{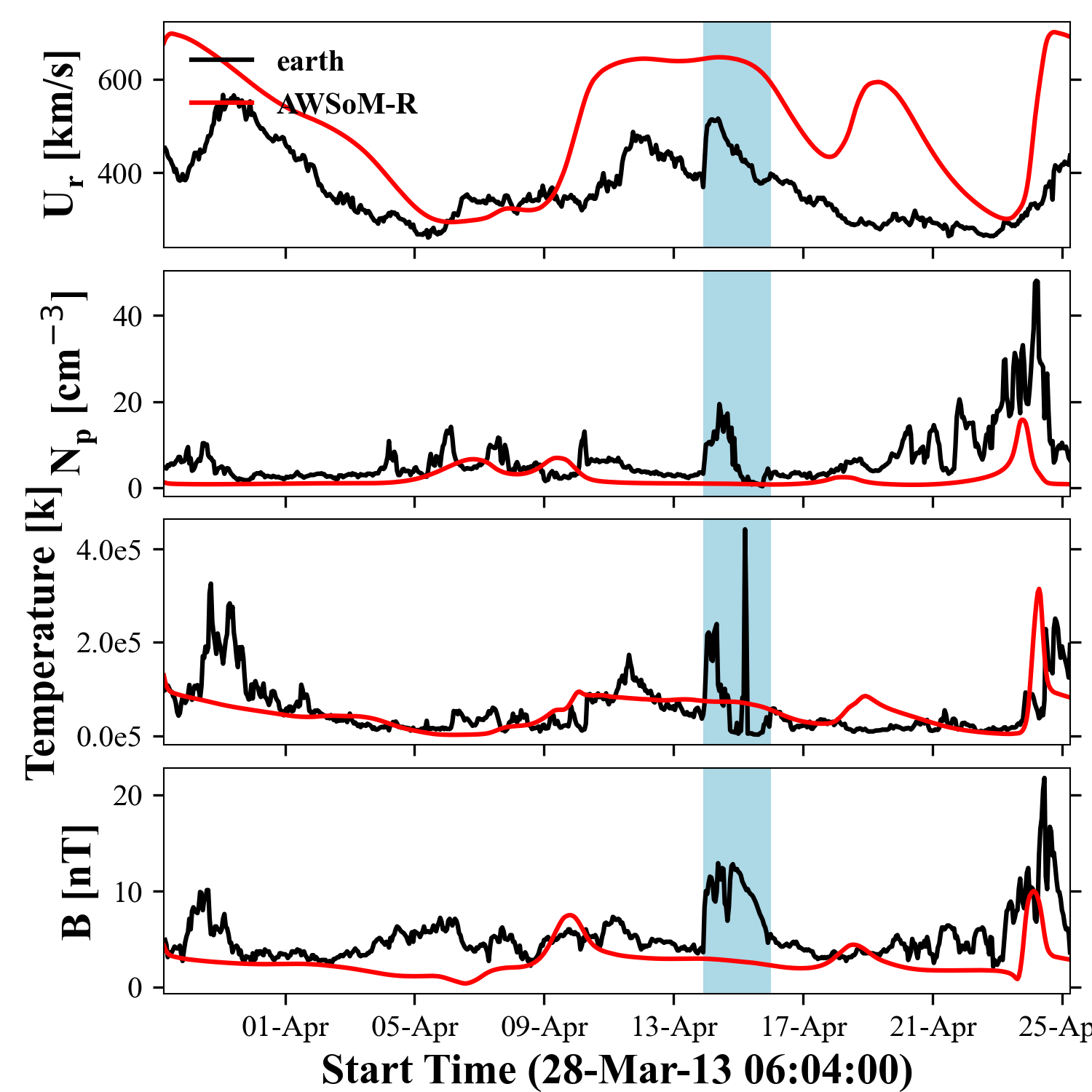}
\noindent\includegraphics[width=0.33\textwidth]{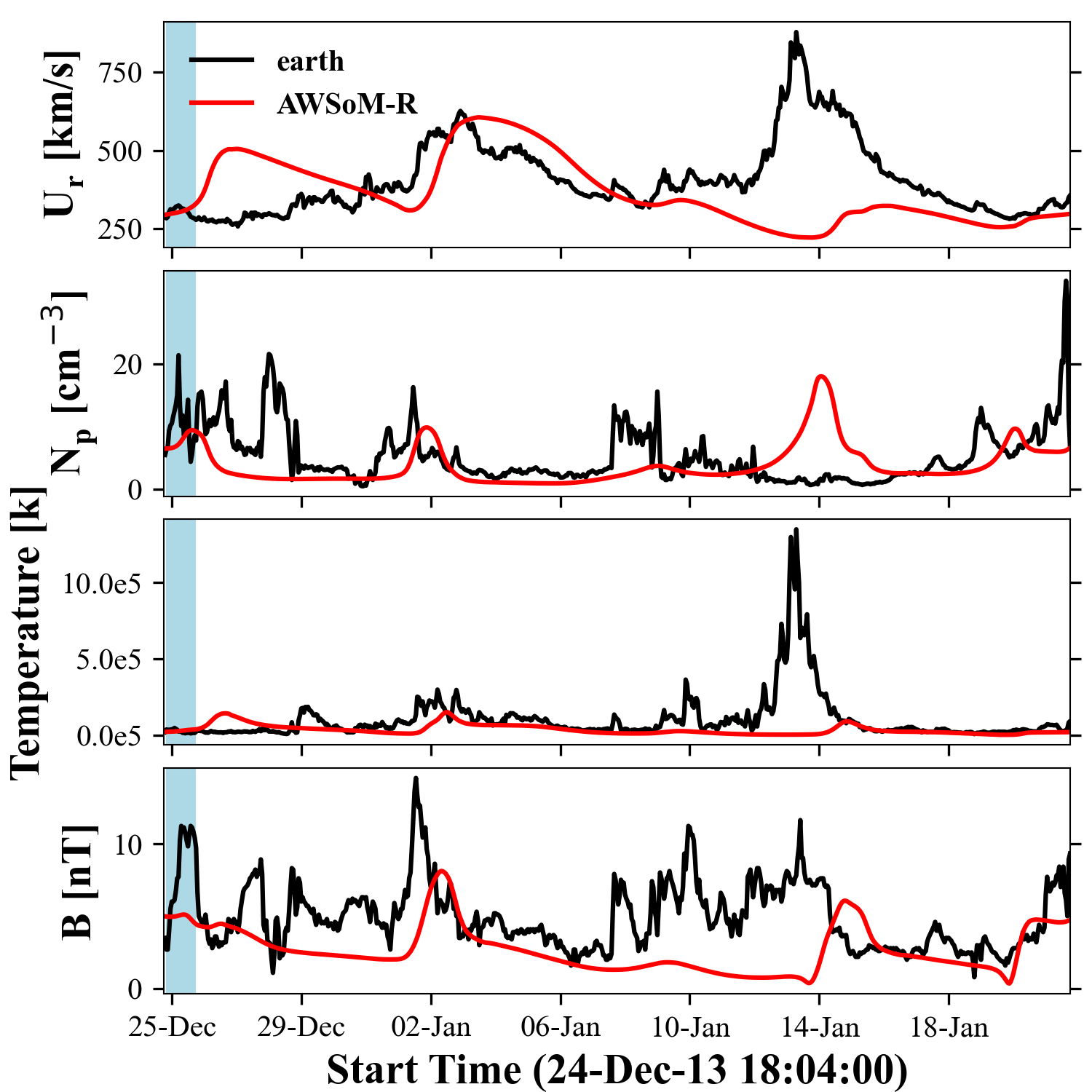}
\noindent\includegraphics[width=0.33\textwidth]{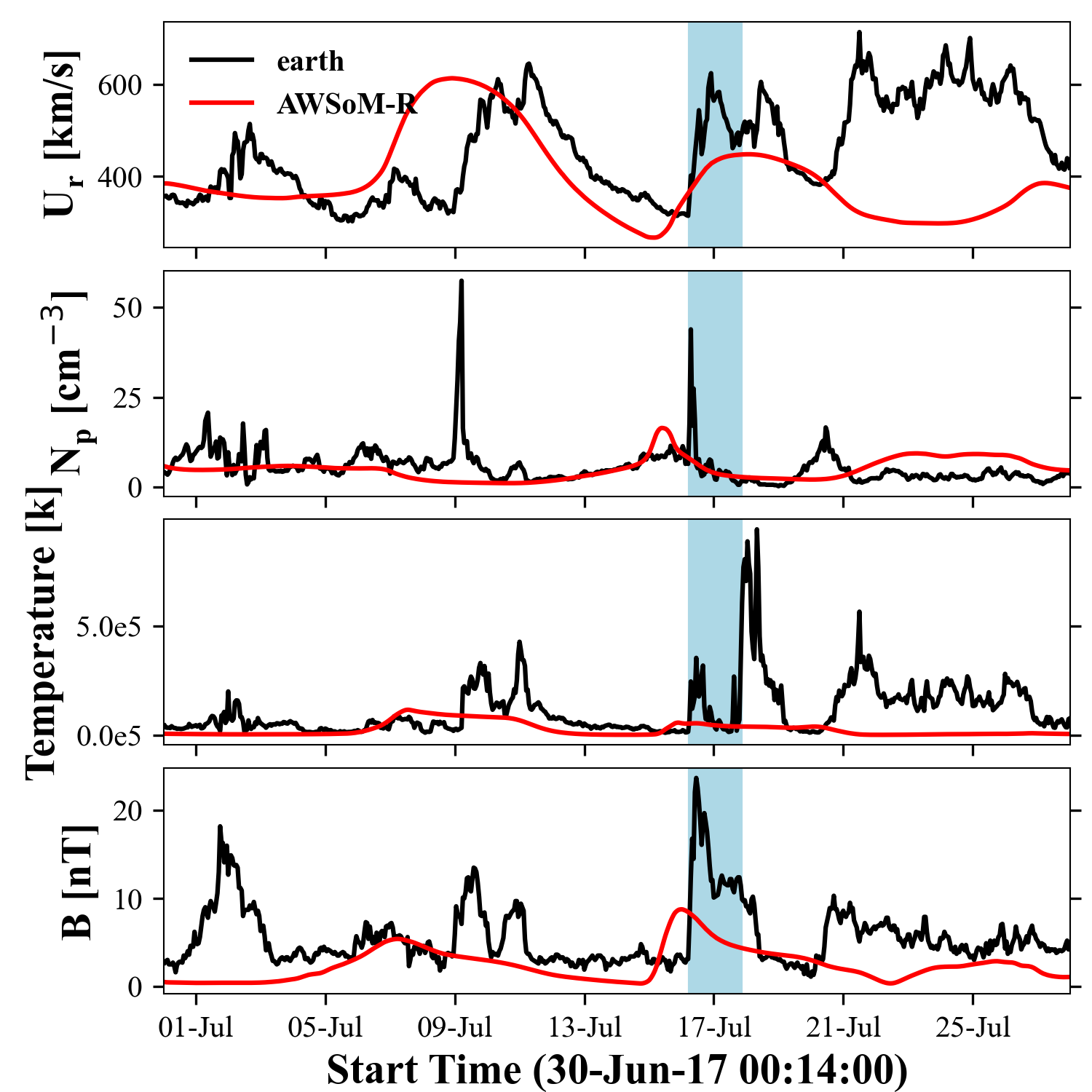}
\noindent\includegraphics[width=0.33\textwidth]{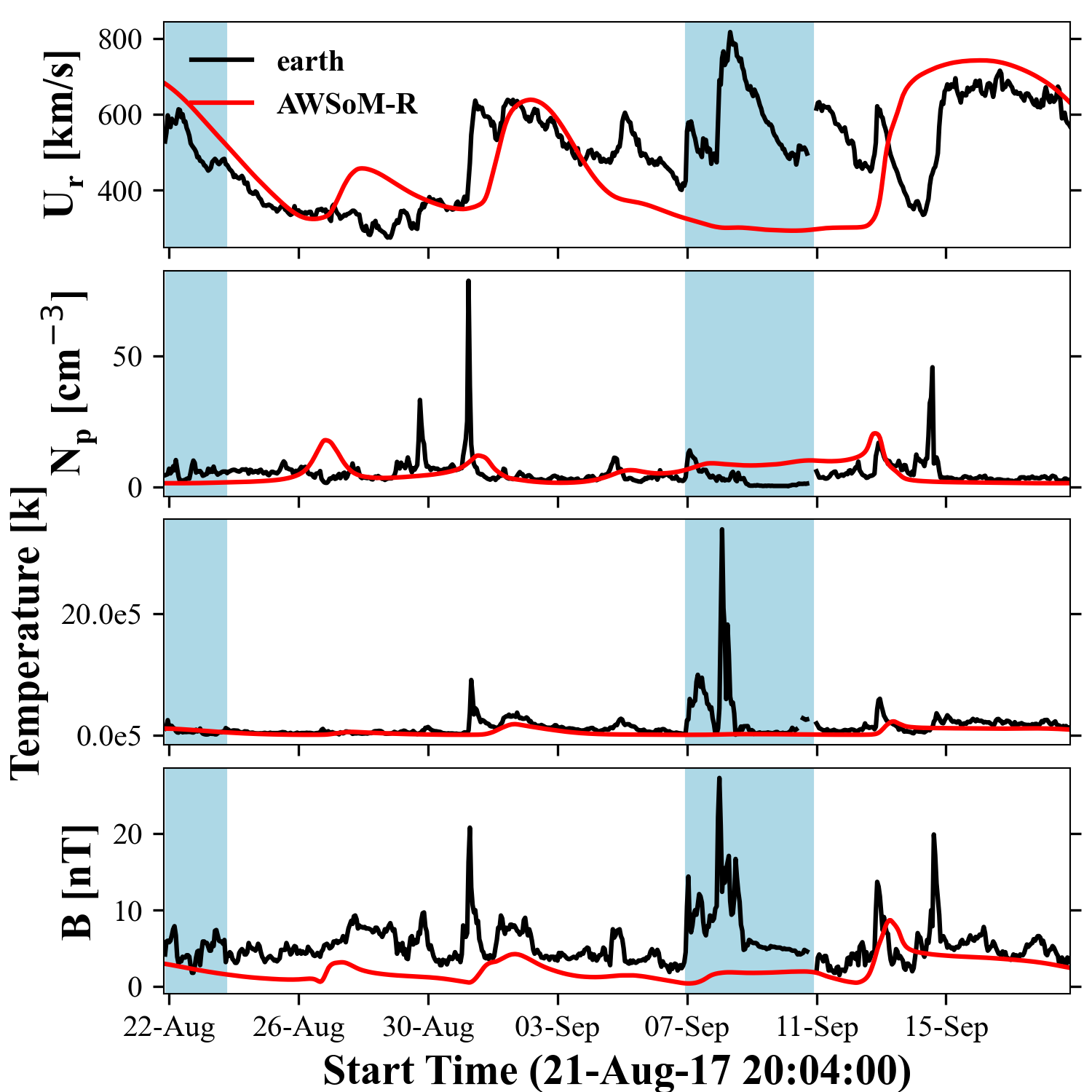}
\noindent\includegraphics[width=0.33\textwidth]{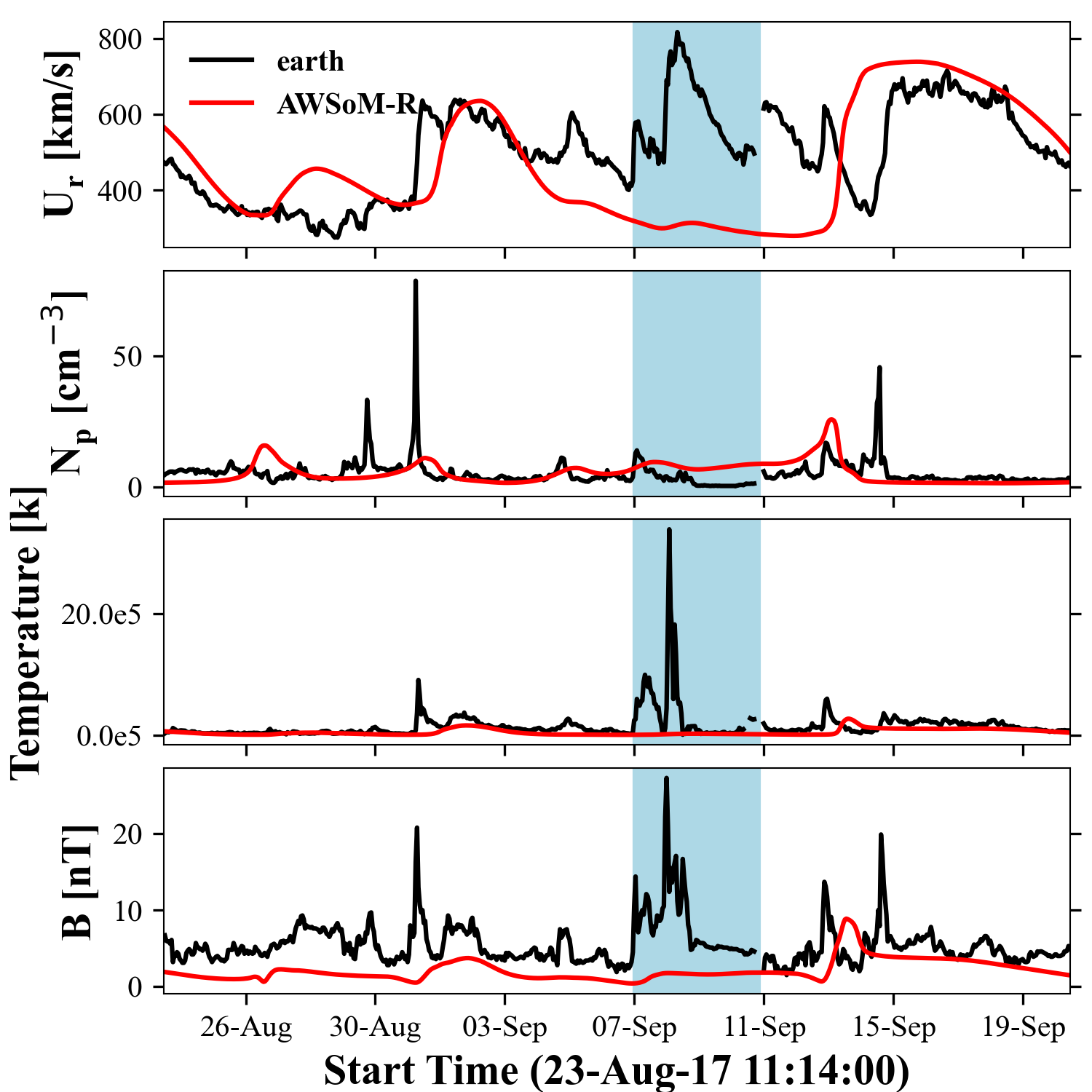}
\noindent\includegraphics[width=0.33\textwidth]{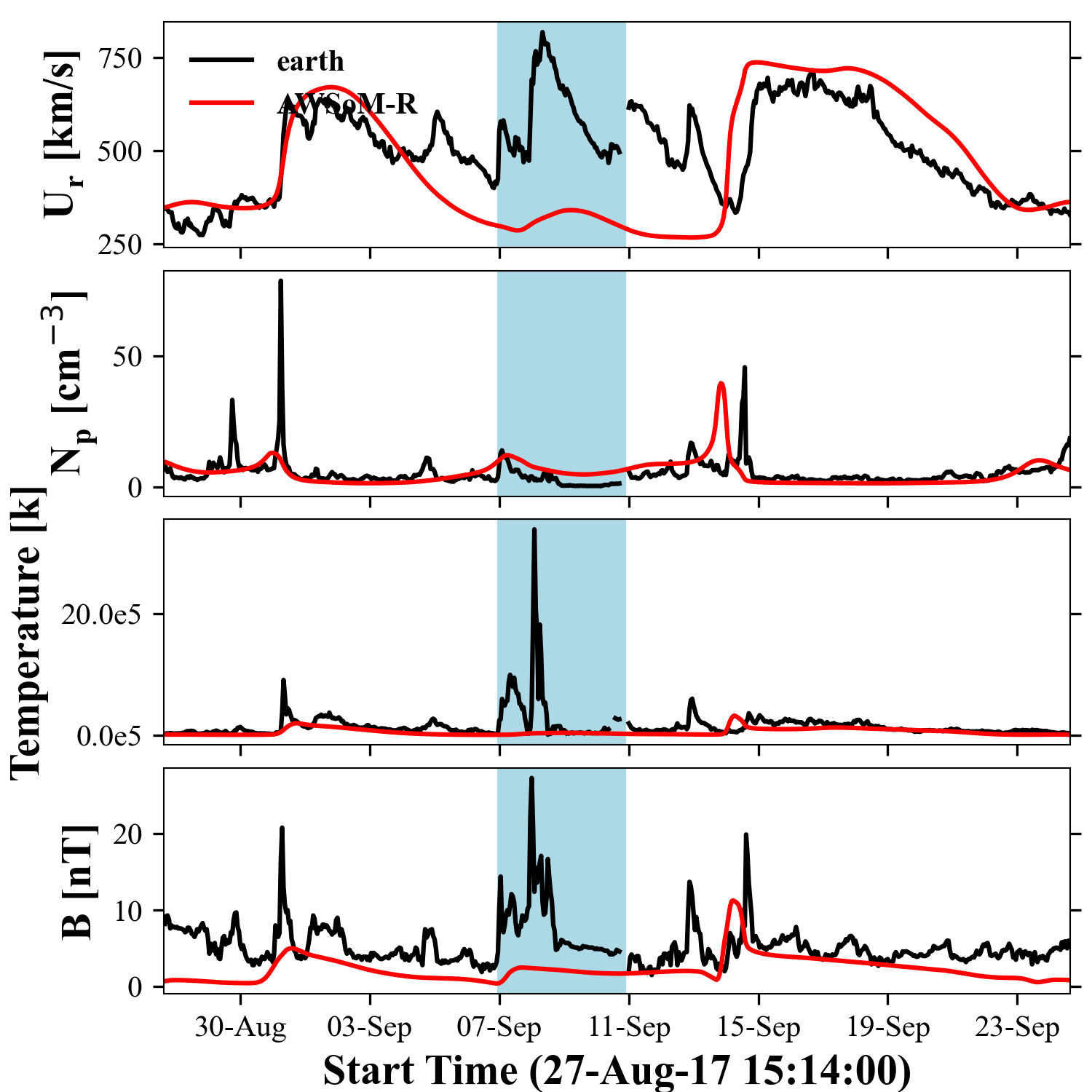}
\caption{Macroscopic properties of the background solar wind for the nine SEP events. In each panel, the radial solar wind plasma speed, the solar wind density, the temperature and the magnitude of the total magnetic field is shown from the top to bottom respectively. The simulation results from AWSoM-R are plotted in red and observations are plotted in black. The passage of the ICME structures are shaded in teal.}
\label{fig:background}
\end{figure}

The highly dynamic solar wind background and the complex geometry of the coronal magnetic field can vary significantly in each Carrington rotation and from event to event.
Therefore, instead of using a homogeneous background solar wind for each event, we calculate the background solar wind properties individually. 
The plasma properties at Earth's location is extracted from the 3D MHD solution and compared with the in-situ measurement made by spacecraft.
As shown in Figure~\ref{fig:background}, the macroscopic properties of the background solar wind for the nine SEP events are shown. For each event, a total time period of $27$ days is shown, corresponding approximately to the synodic solar rotation period. 
In this paper, we only show the in-situ properties of the solar wind and its validation against the observation. The validation of the AWSoM(-R) model using the predicted line-of-sight (LoS) appearance of the corona in different wavelengths has been discussed in detail in \citeA{Sachdeva2019} and \citeA{Gombosi2021}.

In each panel of Figure~\ref{fig:background}, the solar wind properties including the radial bulk plasma speed ($U_r$), the proton number density ($N_p$), the temperature, and the total magnetic field magnitude ($B$) are plotted from top to bottom. 
The simulation results are plotted in red and the observations, measured by the Advanced Composition Explorer, are plotted in black. The time period corresponding to the passage of the ICME are plotted in shaded teal. 
The ICME time periods are obtained from the list of ICMEs observed at 1 AU\footnote{\url{https://izw1.caltech.edu/ACE/ASC/DATA/level3/icmetable2.htm}} \cite{Cane2003,Richardson2010}.
Since we solve the steady state background solar wind, the ICME structures, which are the counterparts of the CMEs in interplanetary space, are not modeled and will not be compared. 
Most of the SEP events occur in solar maximum, especially the ones that we model in this work. Therefore, in multiple panels of Figure~\ref{fig:background}, one can see more than one ICMEs in the observations. As we mentioned above, the ICMEs in the observations will not be captured by the simulation. The mismatch between the simulation and observation in the ICME time period is as expected.
Except the ICMEs, the overall dynamics of the solar wind plasma are well-represented by the simulation. 

When running the AWSoM-R model, to get a reasonable comparison between the simulations and observations, there are 
two adjustable input parameters: the Poynting flux parameter and the correlation length of the Alfv\'en wave dissipation
\cite<see details in>{Huang2023,vanderHolst2014,Jivani2023}.
The Poynting flux parameter determines the input energy at the inner boundary to heat the solar corona and accelerate the solar wind, 
and the correlation length describes the dissipation of Alfv\'en wave turbulence in the solar corona and heliosphere \cite{Huang2023}.
When running the AWSoM-R model to obtain the background solar wind, we varied the Poynting Flux parameter to get the best comparison between the simulations and observations.
A detailed discussion on choosing the optimal Poynting flux parameter is discussed in detail in a recent paper by \citeA{Huang2023}.

\subsection{CMEs}\label{sec:cme}

\begin{figure}
\noindent\includegraphics[width=0.33\textwidth]{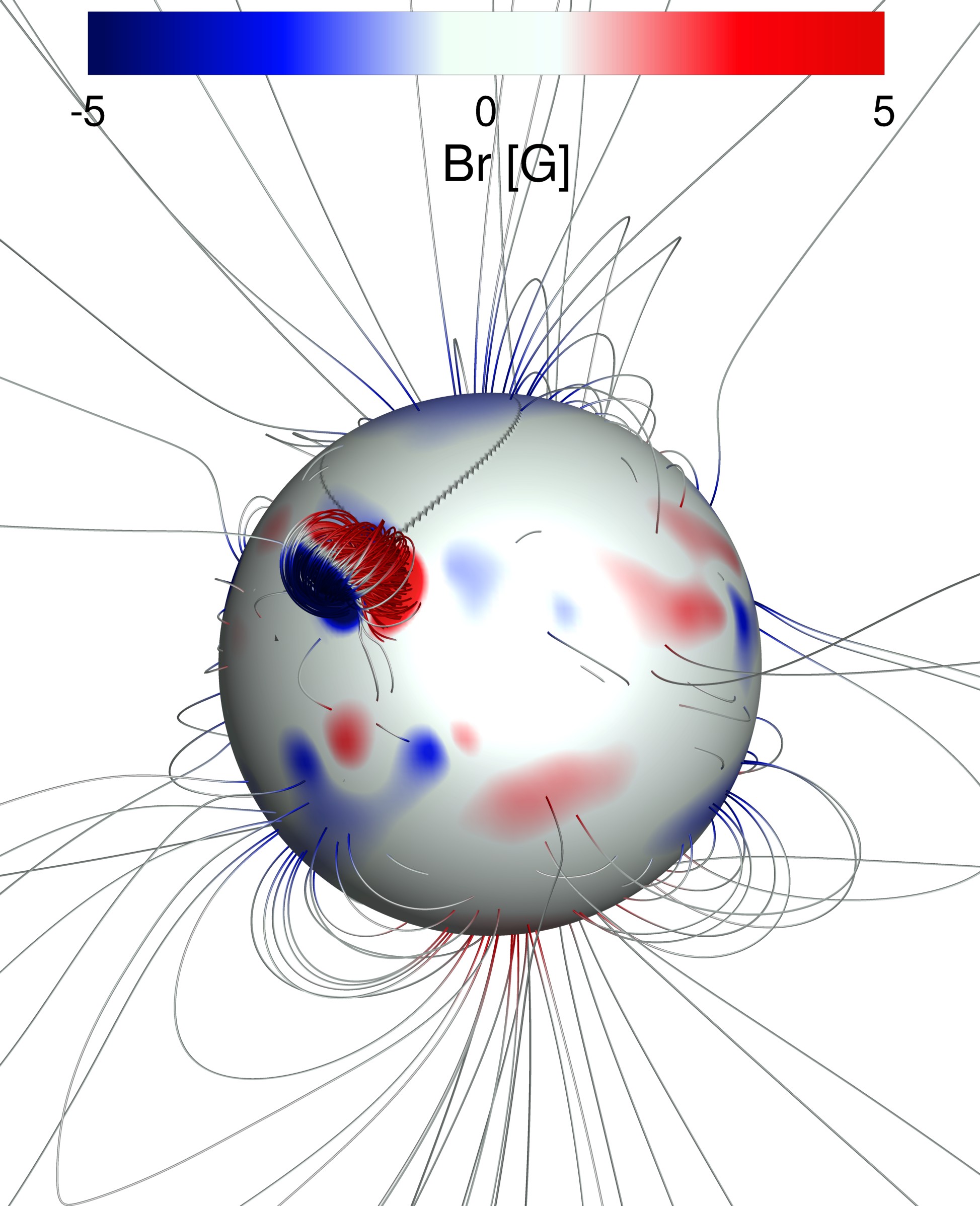}
\noindent\includegraphics[width=0.33\textwidth]{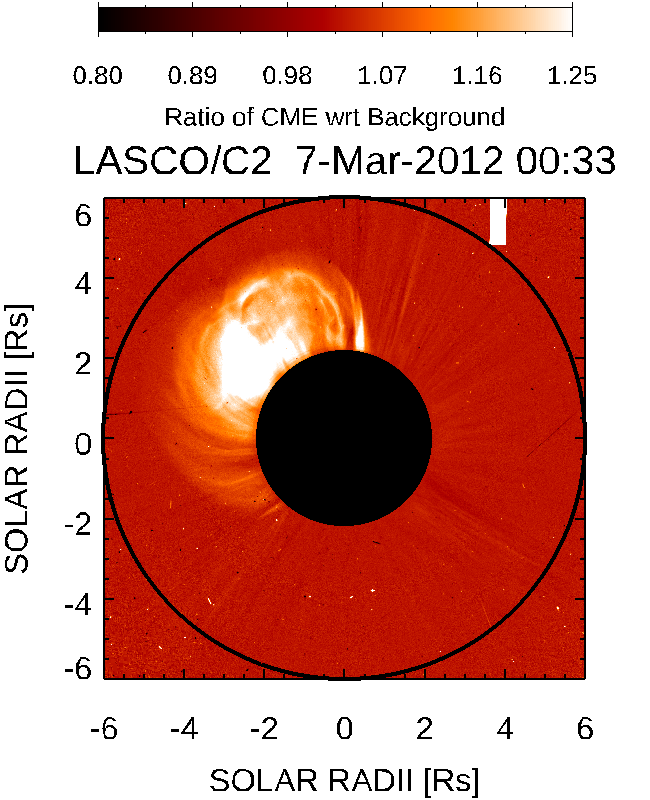}
\noindent\includegraphics[width=0.33\textwidth]{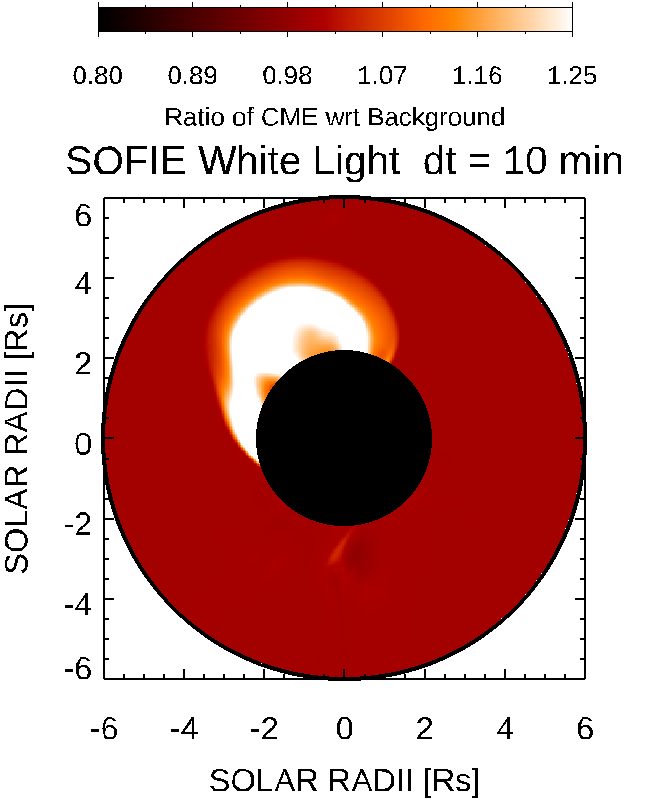}
\noindent\includegraphics[width=0.33\textwidth]{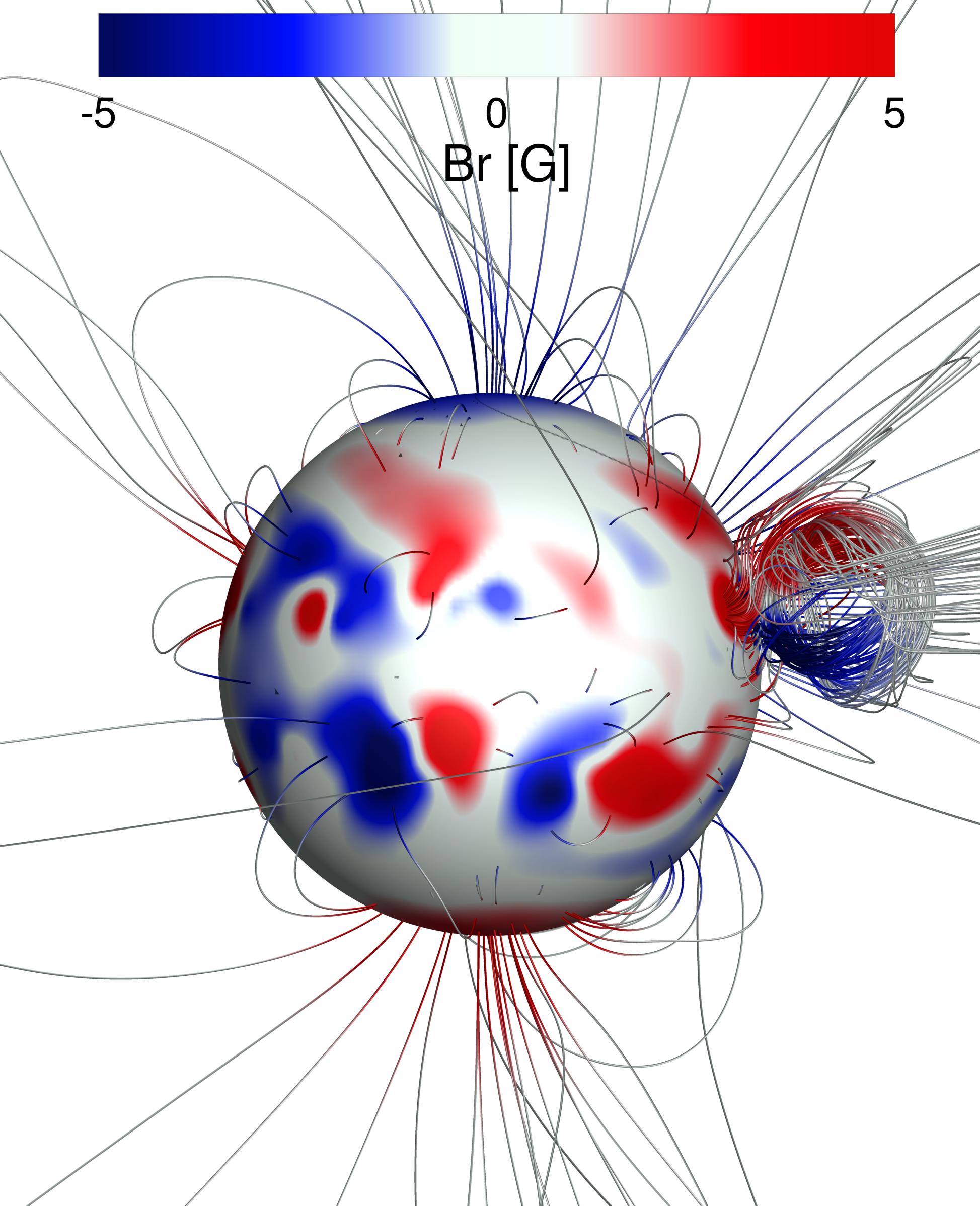}
\noindent\includegraphics[width=0.33\textwidth]{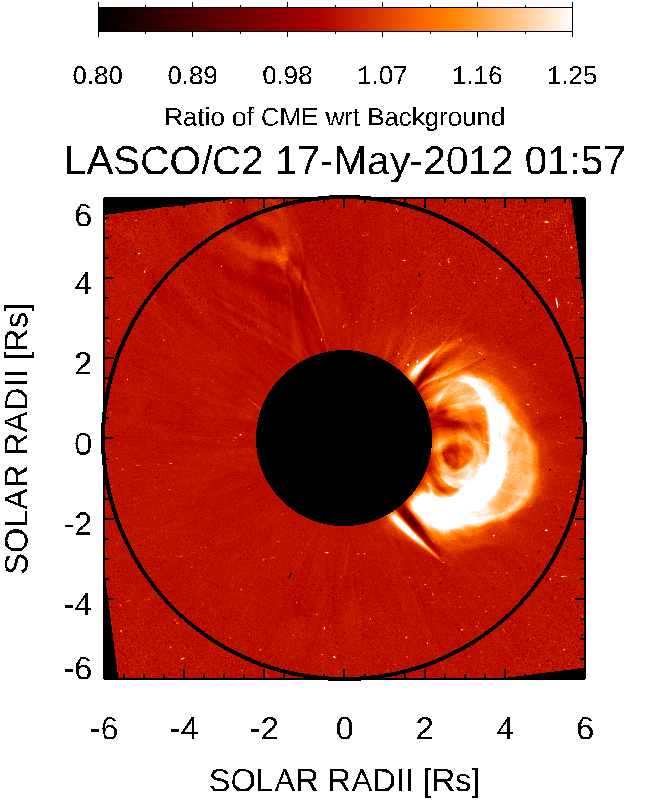}
\noindent\includegraphics[width=0.33\textwidth]{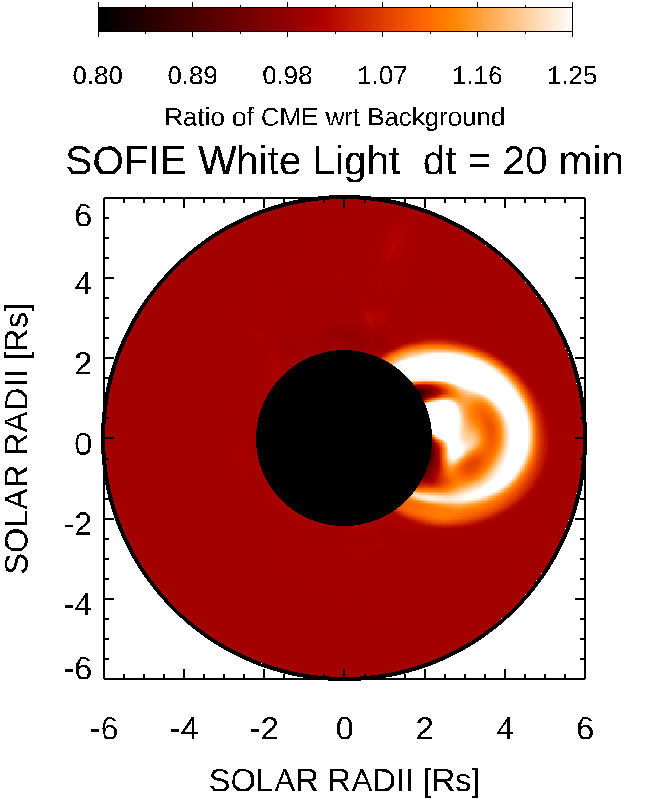}
\noindent\includegraphics[width=0.33\textwidth]{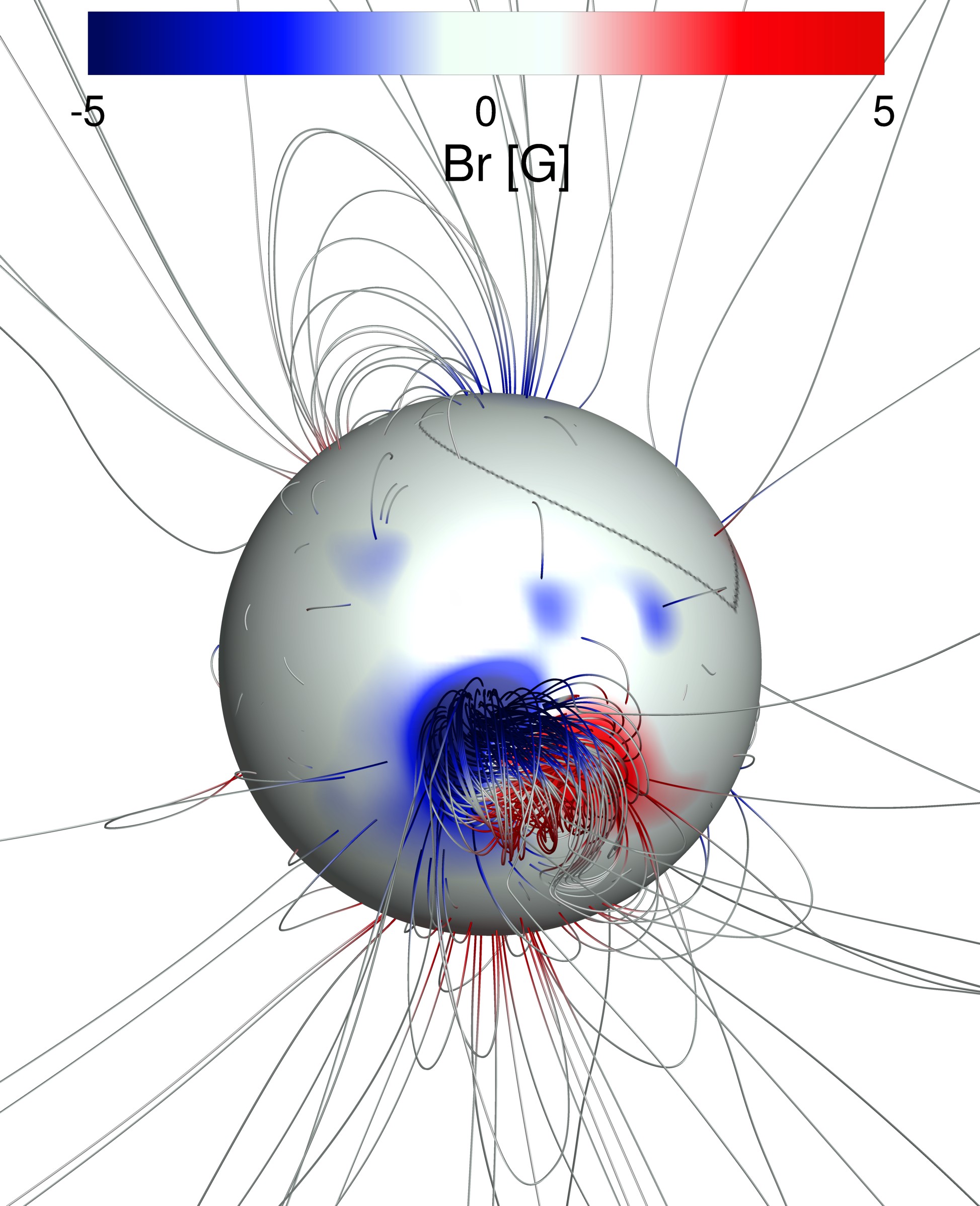}
\noindent\includegraphics[width=0.33\textwidth]{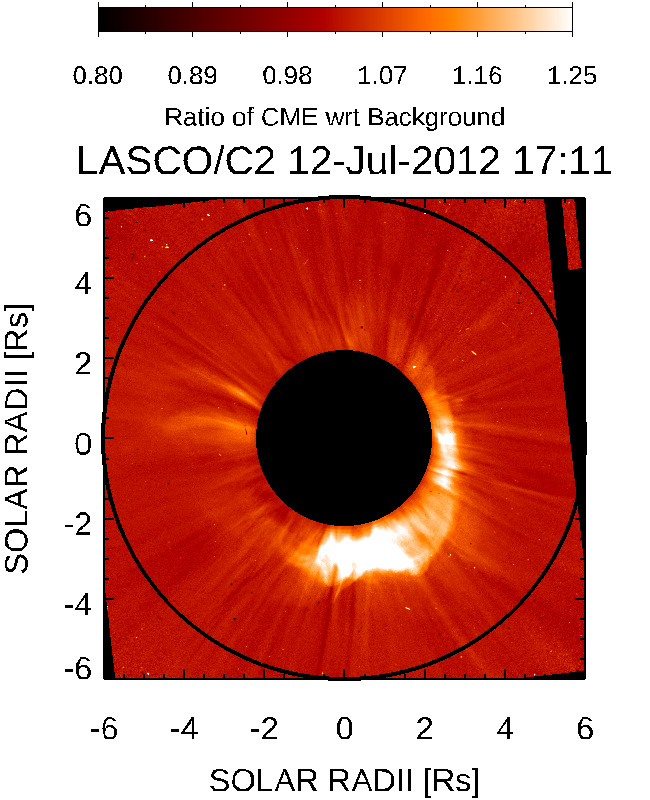}
\noindent\includegraphics[width=0.33\textwidth]{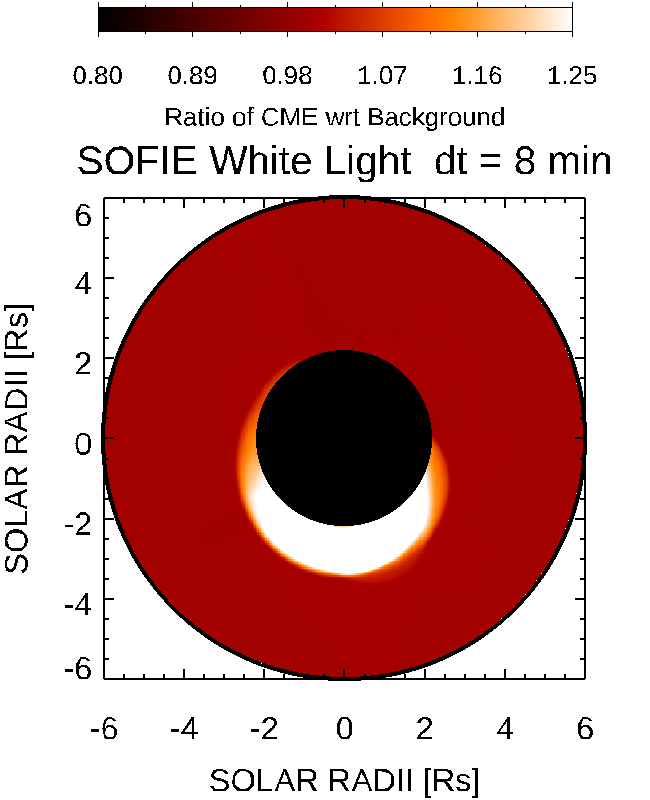}
\caption{Left: The 3D topology of the inserted magnetic flux rope in the active region. Middle: LASCO/C2 white-light image of the solar corona. Right: White-light image calculated from the simulation at the same time as the middle column. Three events are shown here, 2012-Mar-07, 2012-May-17, and 2012-Jul-12. In the left column, the surface of the Sun (1.105 Rs) and the 3D magnetic field lines are colored with the radial magnetic field. The color bar shown in the plot presents the strength of the radial magnetic field in the field lines. The radial magnetic field on the Sun ranges from $-20$ Gauss to $20$ Gauss (color bar not shown here).}
\label{fig:cme_1}
\end{figure}

\begin{figure}
\noindent\includegraphics[width=0.33\textwidth]{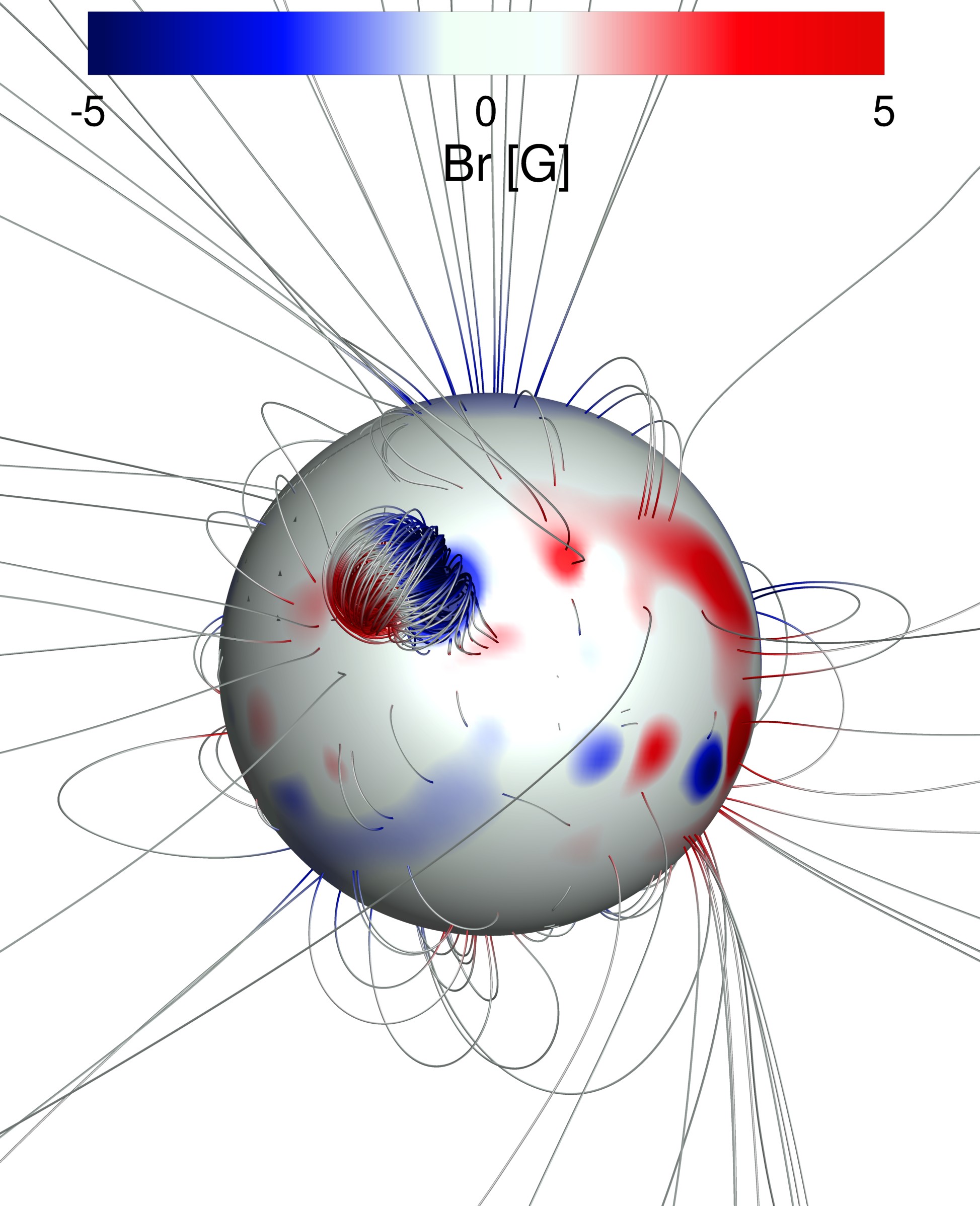}
\noindent\includegraphics[width=0.33\textwidth]{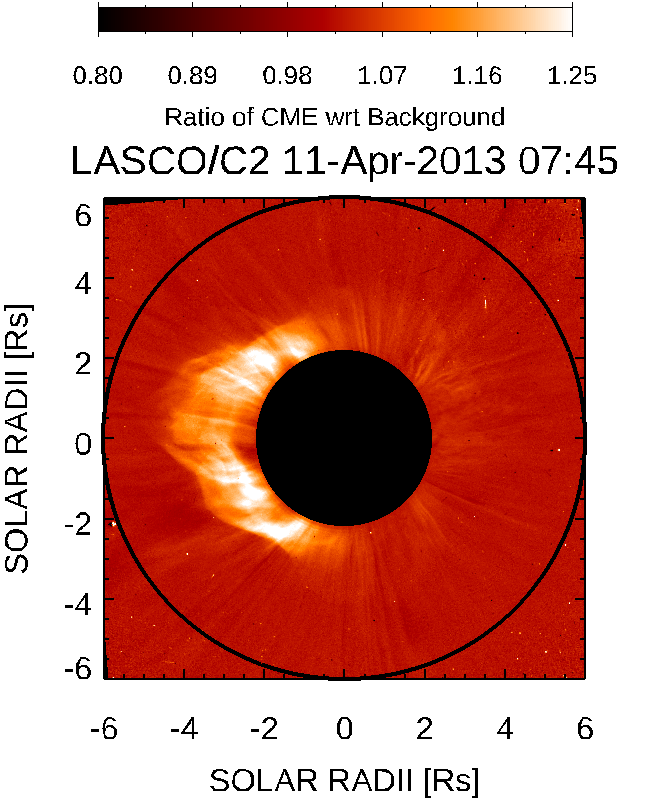}
\noindent\includegraphics[width=0.33\textwidth]{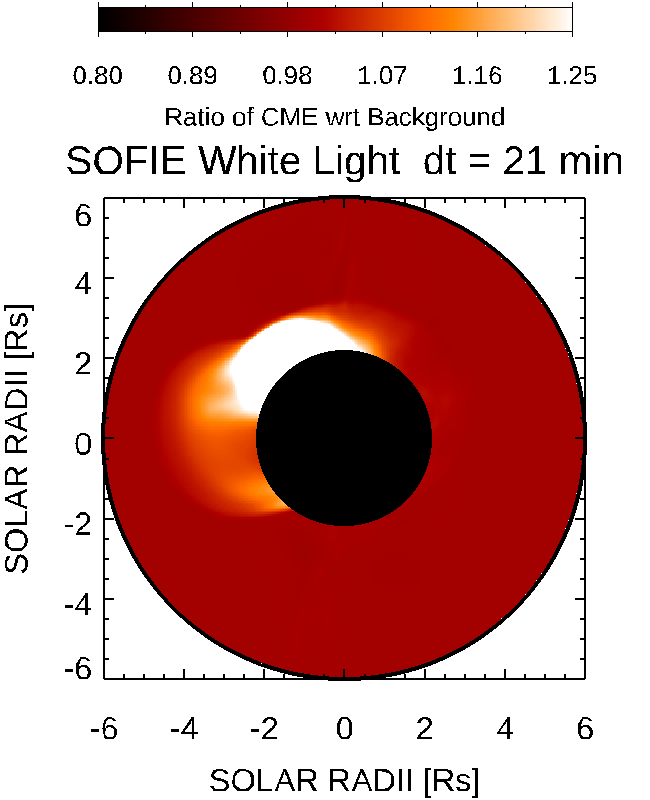}
\noindent\includegraphics[width=0.33\textwidth]{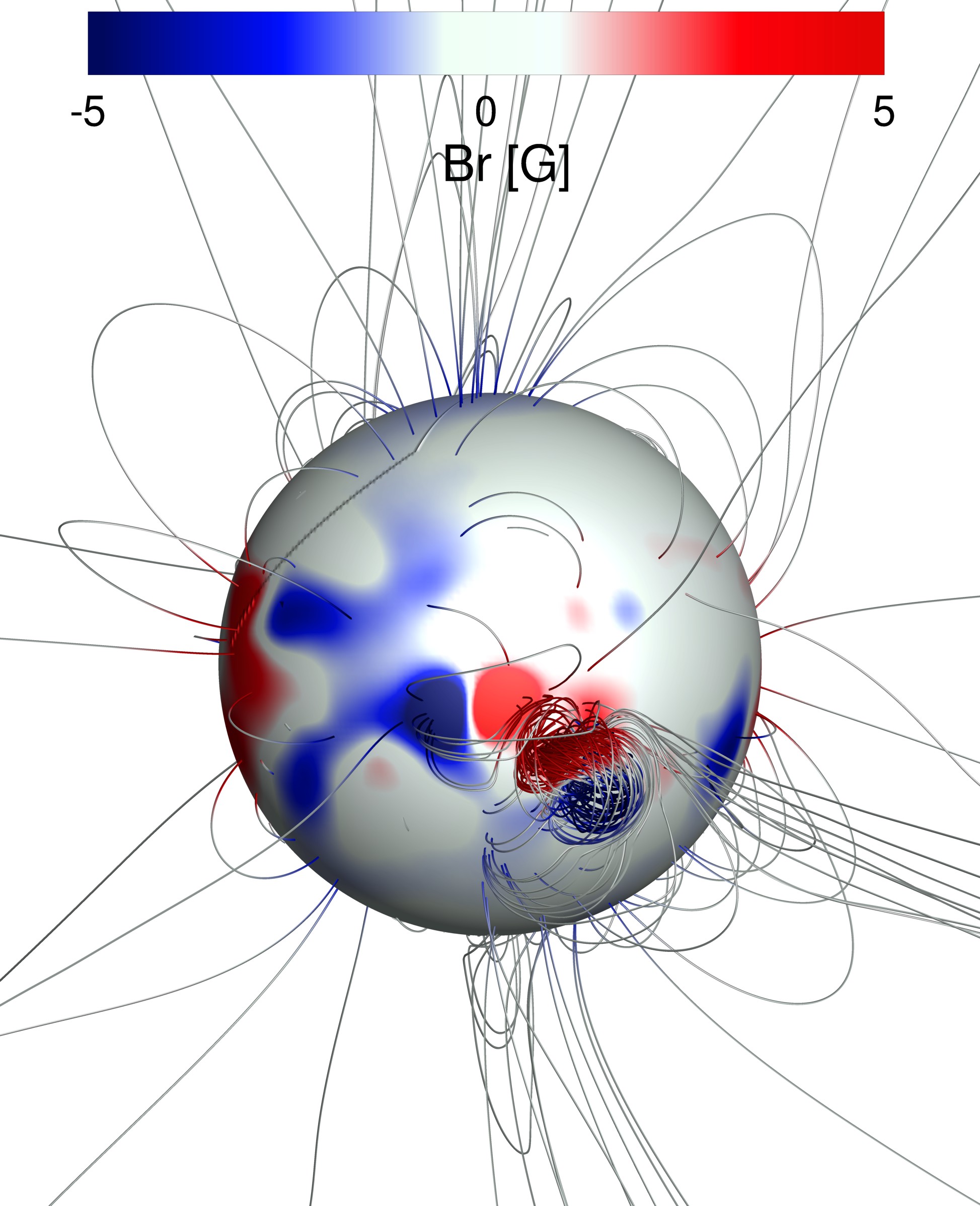}
\noindent\includegraphics[width=0.33\textwidth]{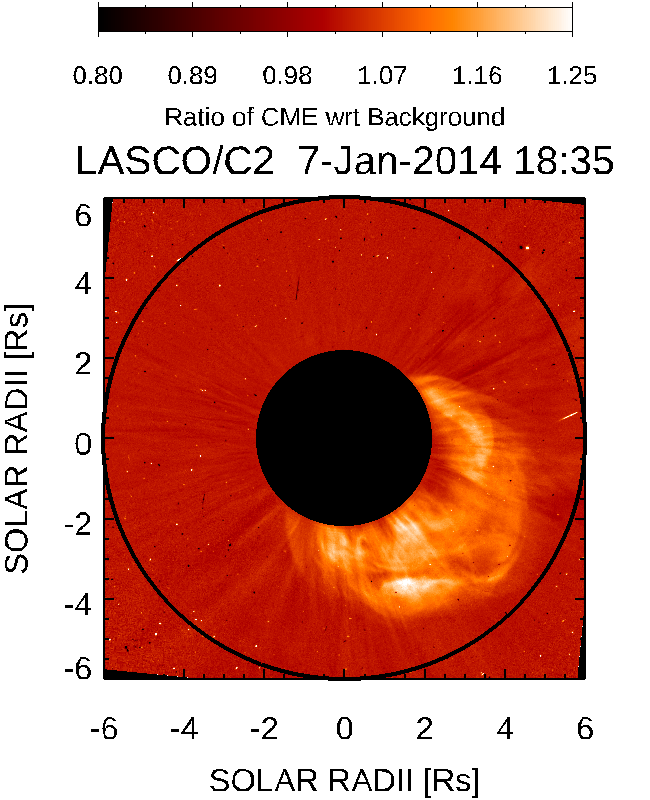}
\noindent\includegraphics[width=0.33\textwidth]{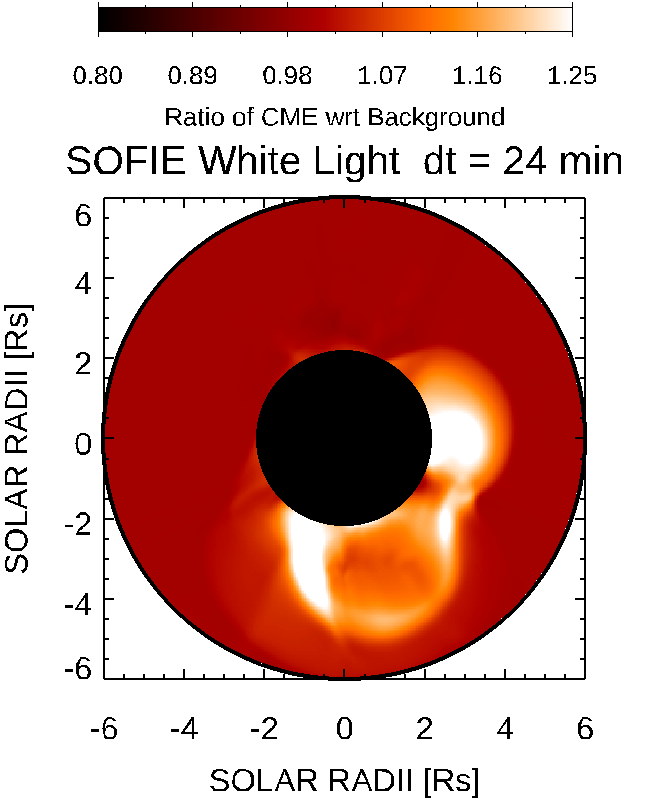}
\noindent\includegraphics[width=0.33\textwidth]{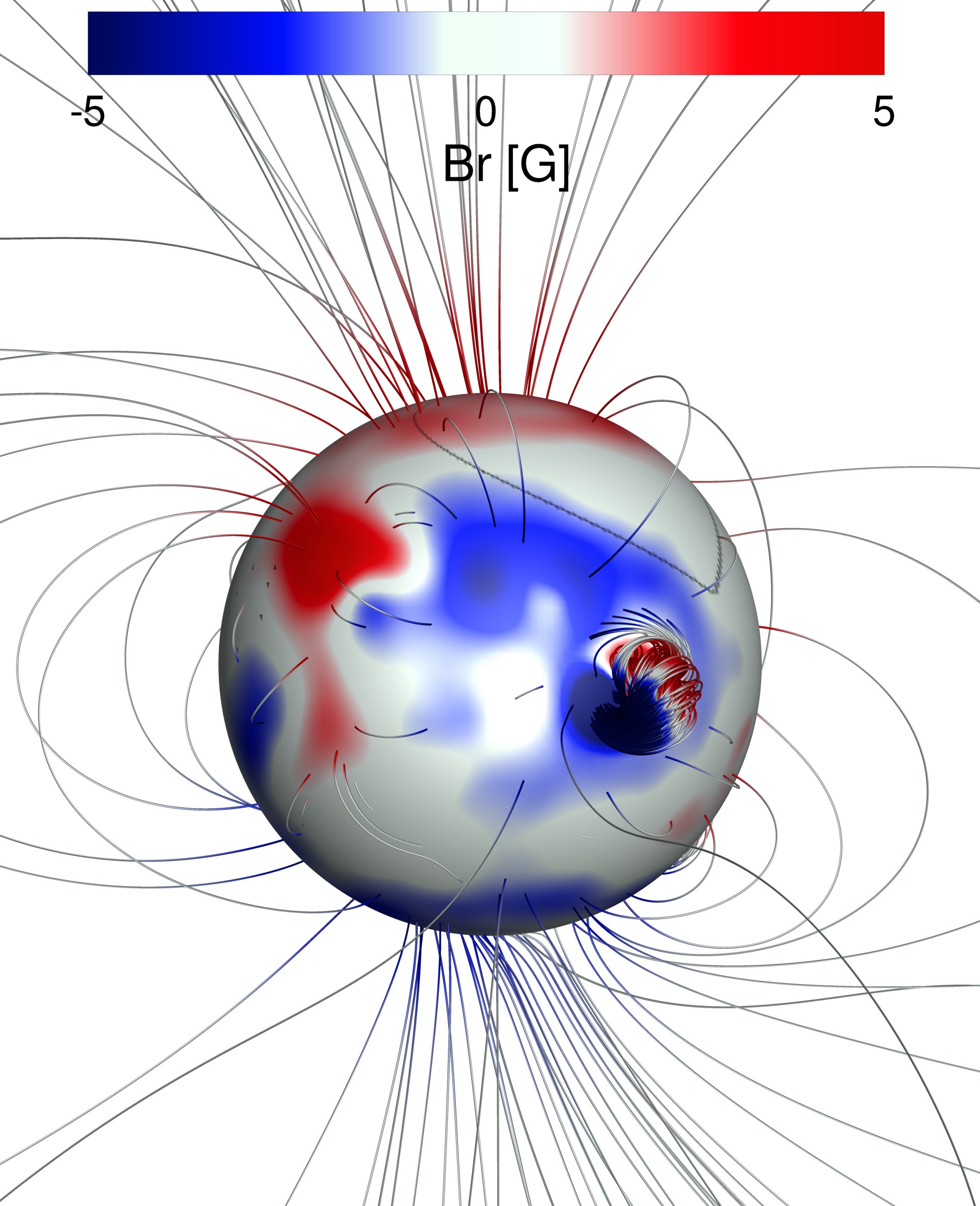}
\noindent\includegraphics[width=0.33\textwidth]{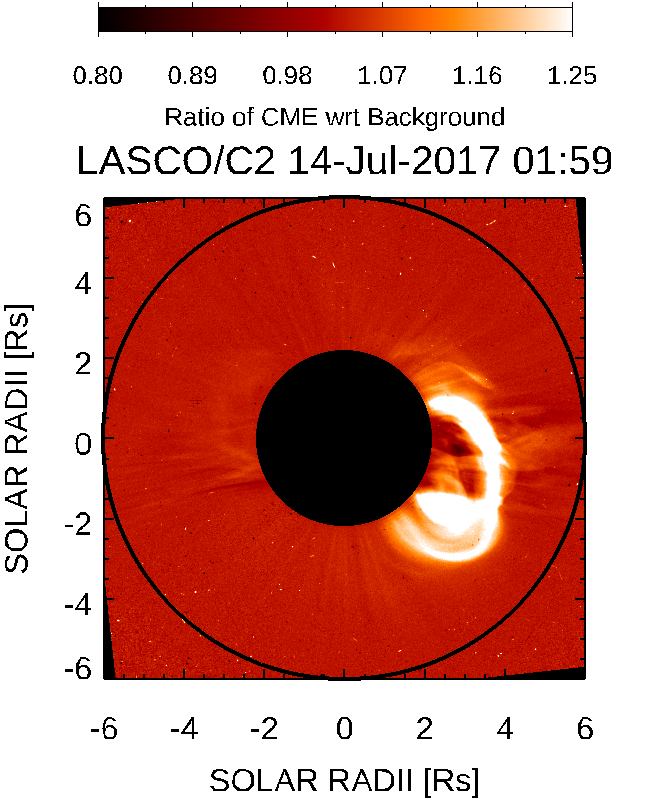}
\noindent\includegraphics[width=0.33\textwidth]{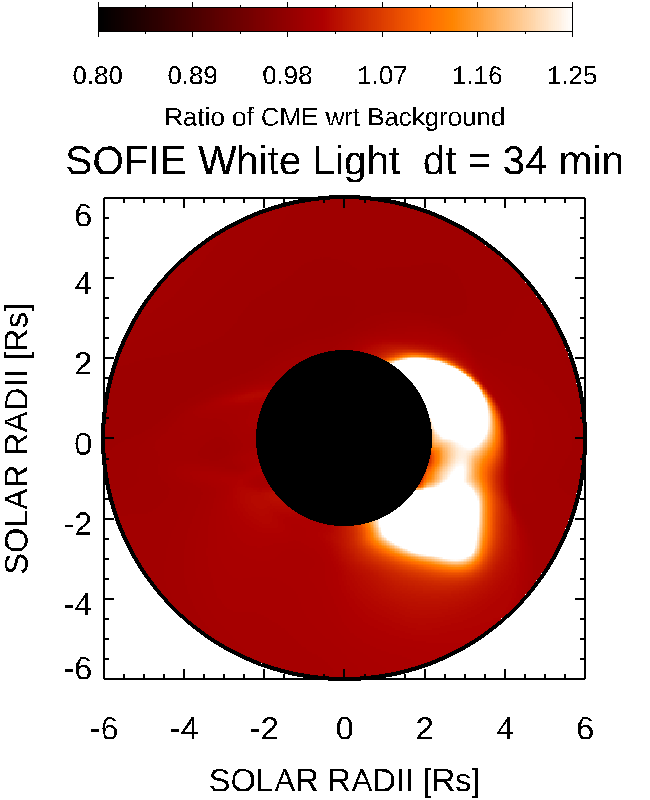}
\caption{In the same format as Figure~\ref{fig:cme_1} for the three events 2013-Apr-11, 2014-Jan-07, and 2017-Jul-14.}
\label{fig:cme_2}
\end{figure}

\begin{figure}
\noindent\includegraphics[width=0.33\textwidth]{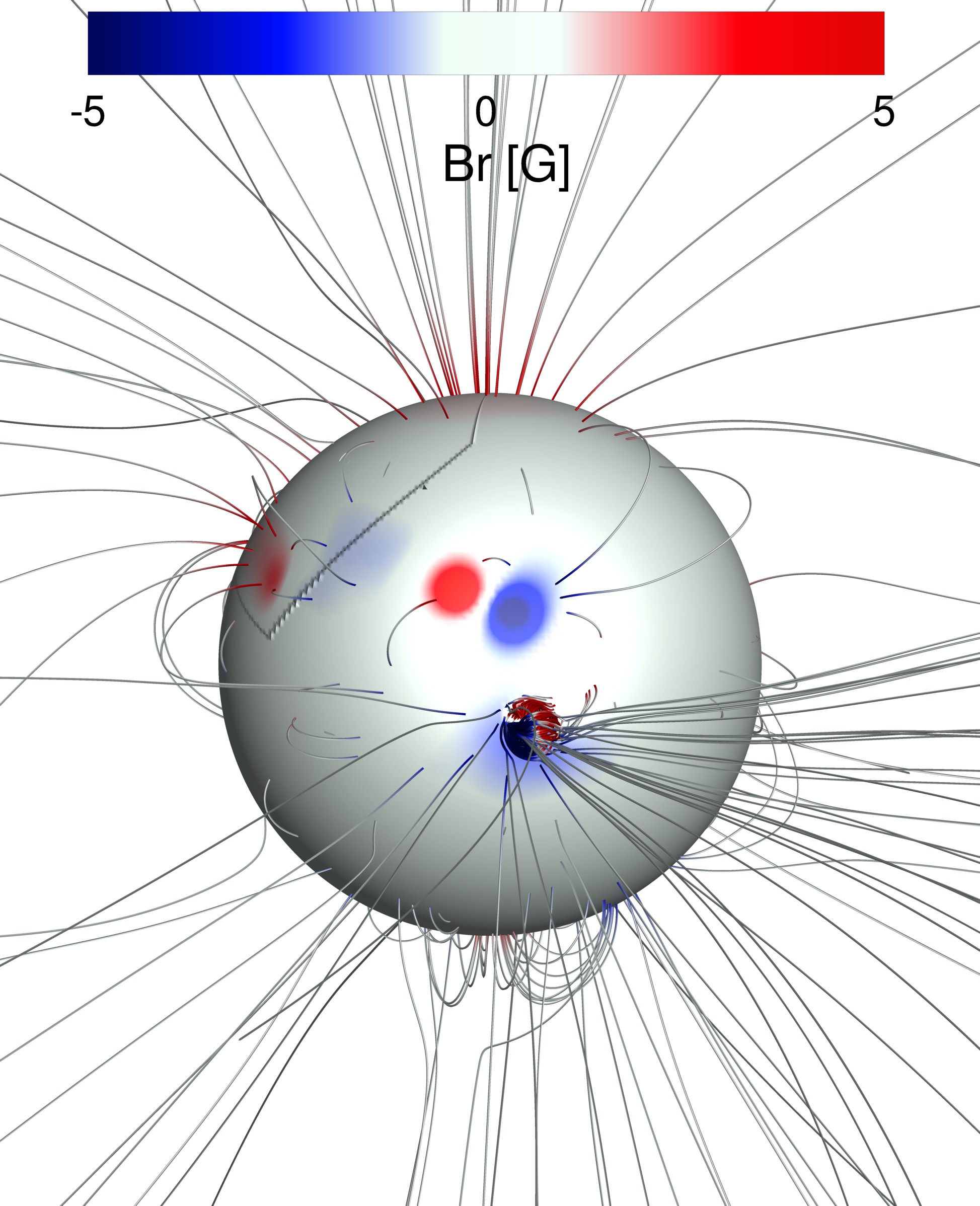}
\noindent\includegraphics[width=0.33\textwidth]{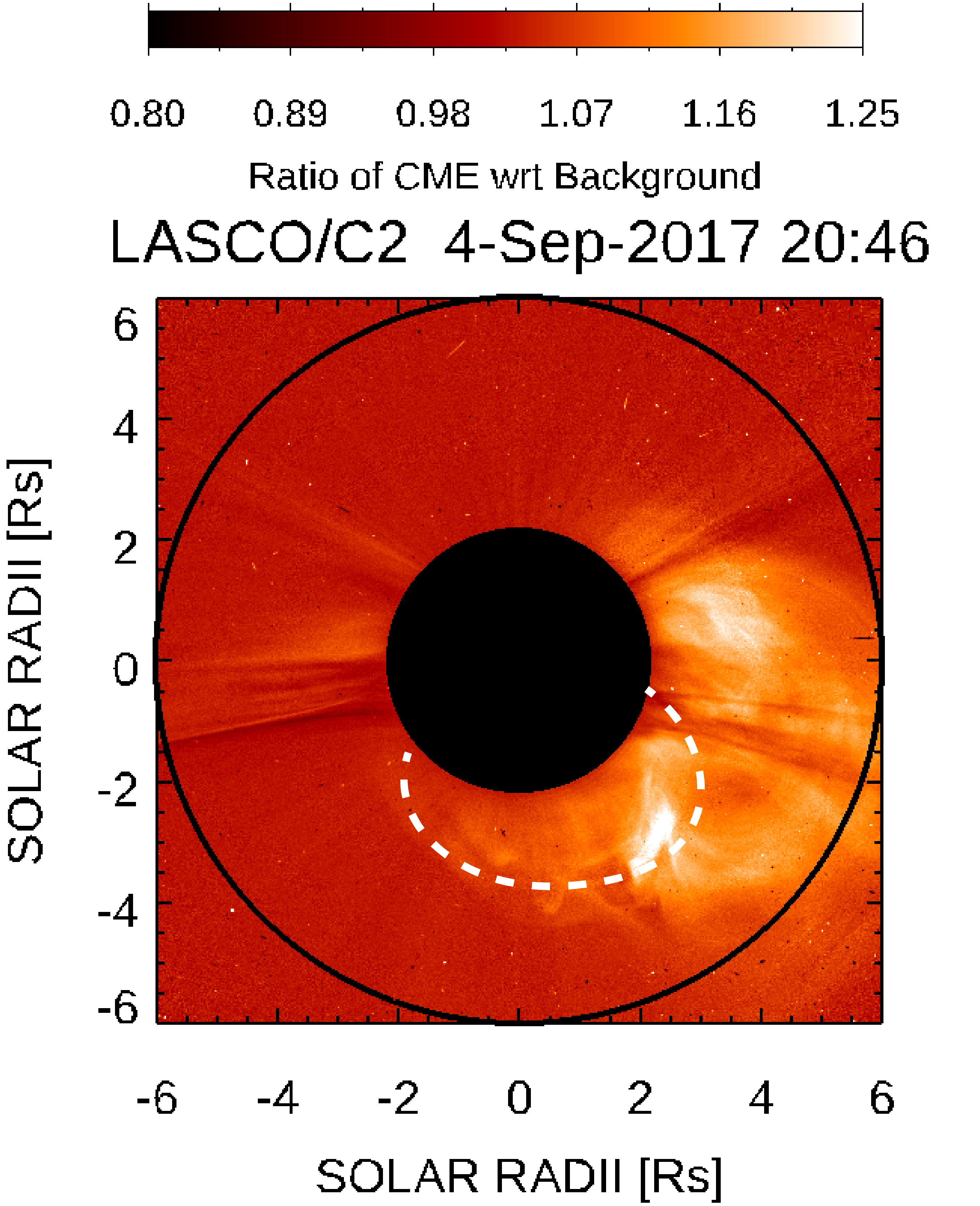}
\noindent\includegraphics[width=0.33\textwidth]{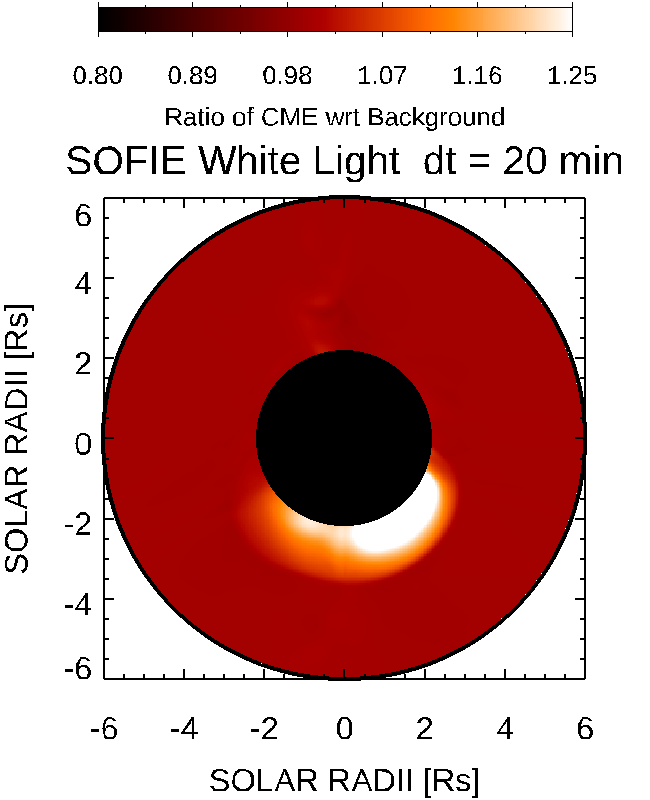}
\noindent\includegraphics[width=0.33\textwidth]{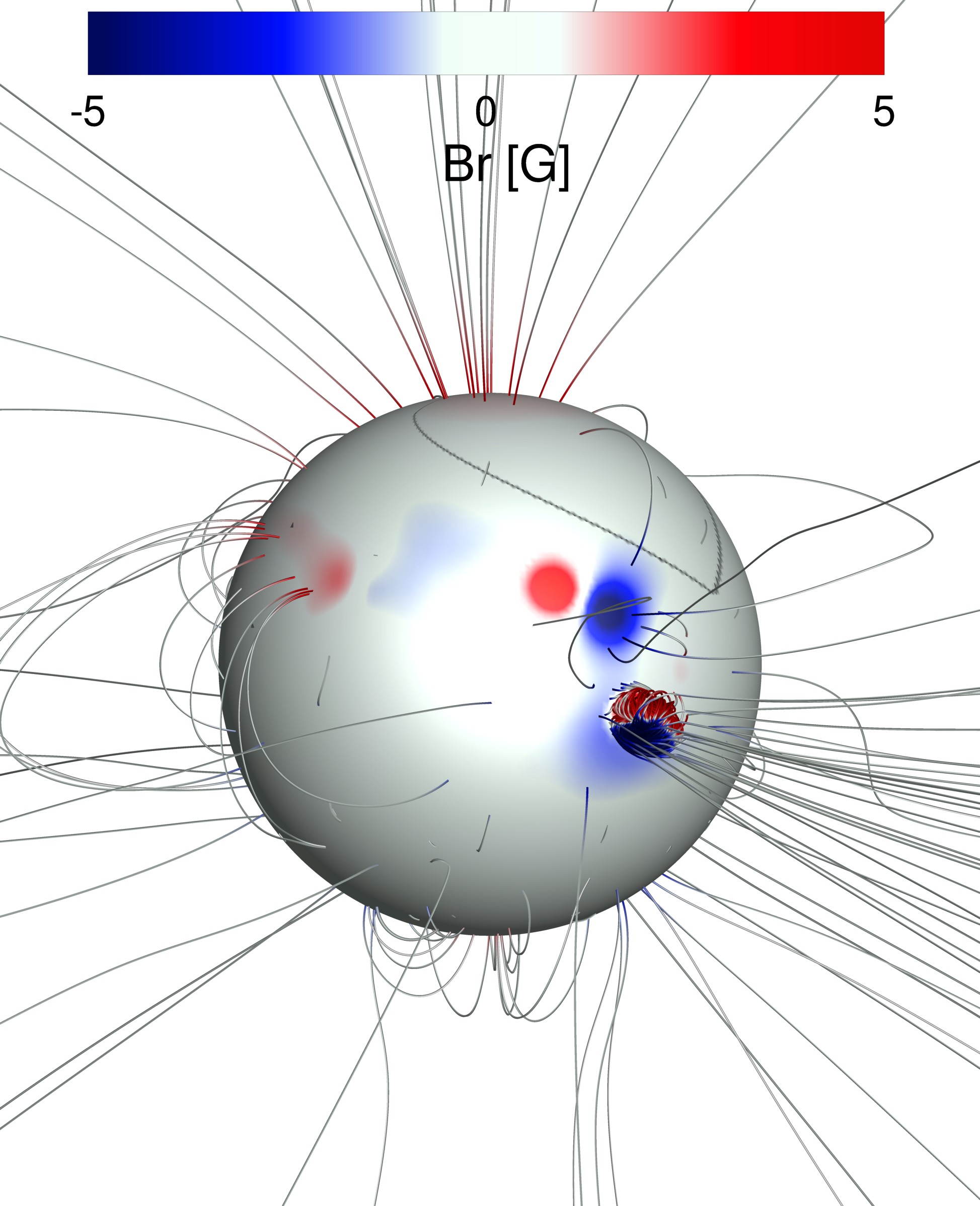}
\noindent\includegraphics[width=0.33\textwidth]{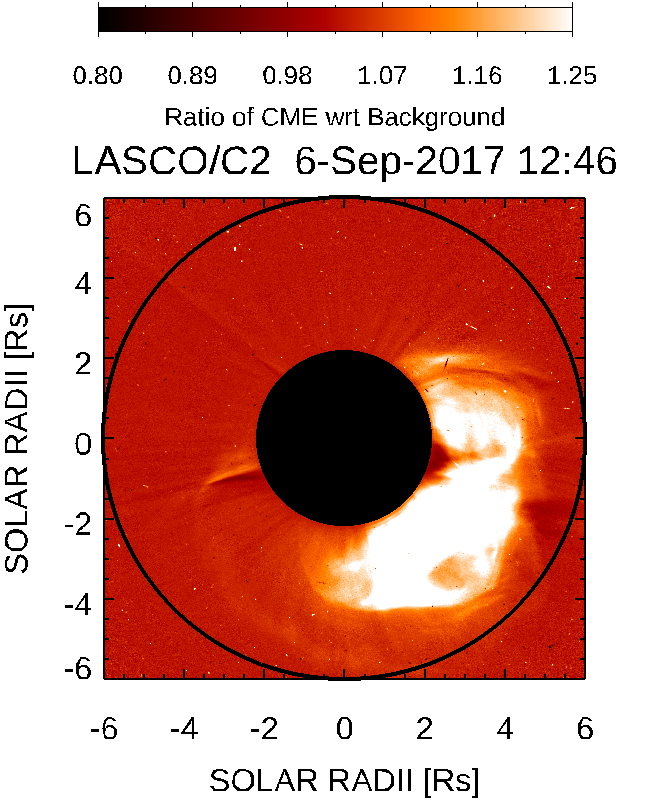}
\noindent\includegraphics[width=0.33\textwidth]{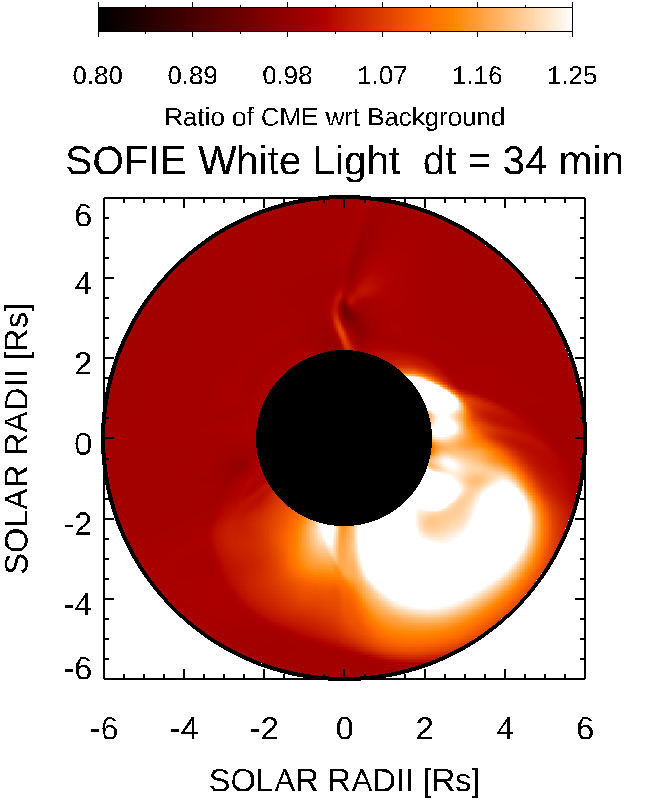}
\noindent\includegraphics[width=0.33\textwidth]{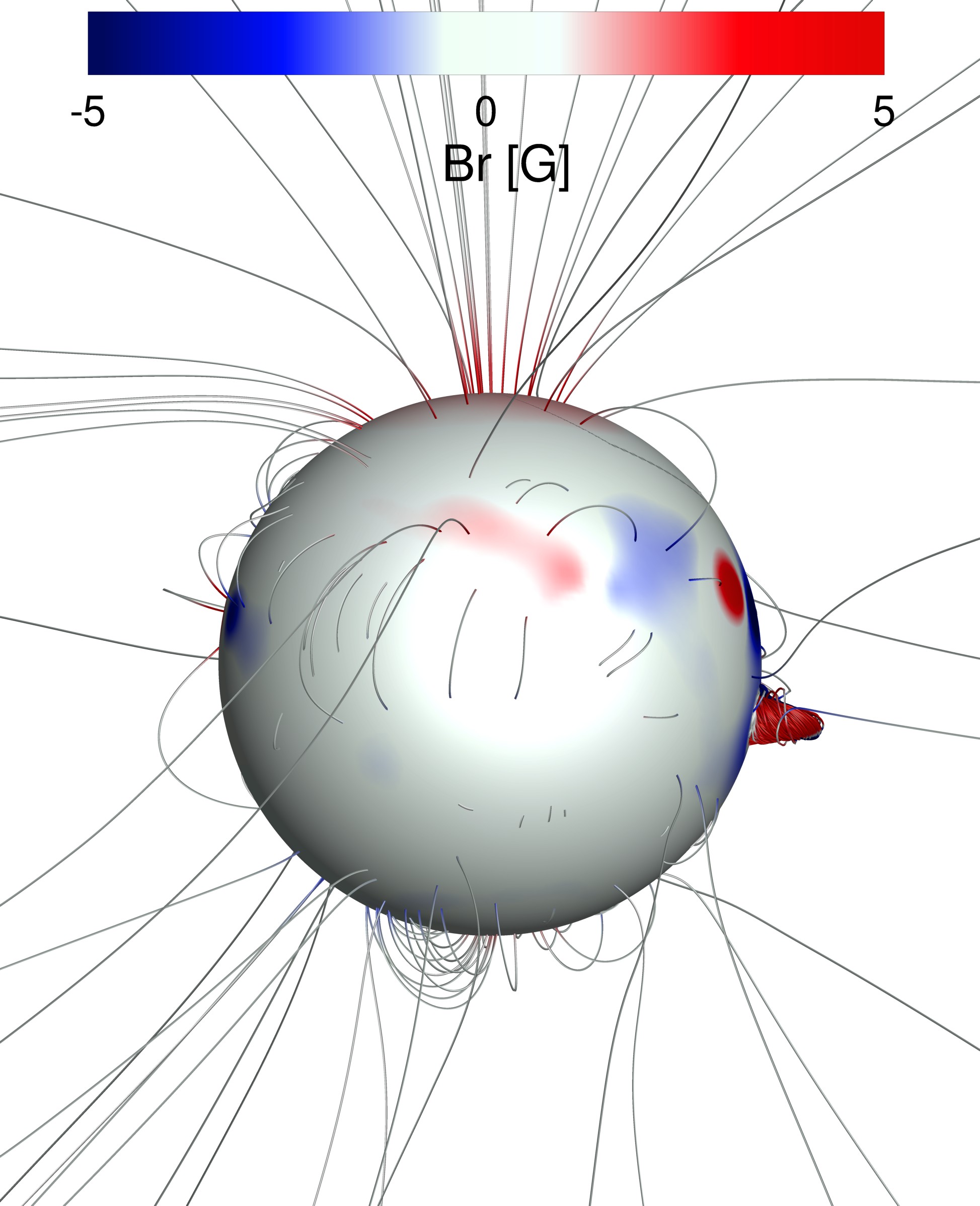}
\noindent\includegraphics[width=0.33\textwidth]{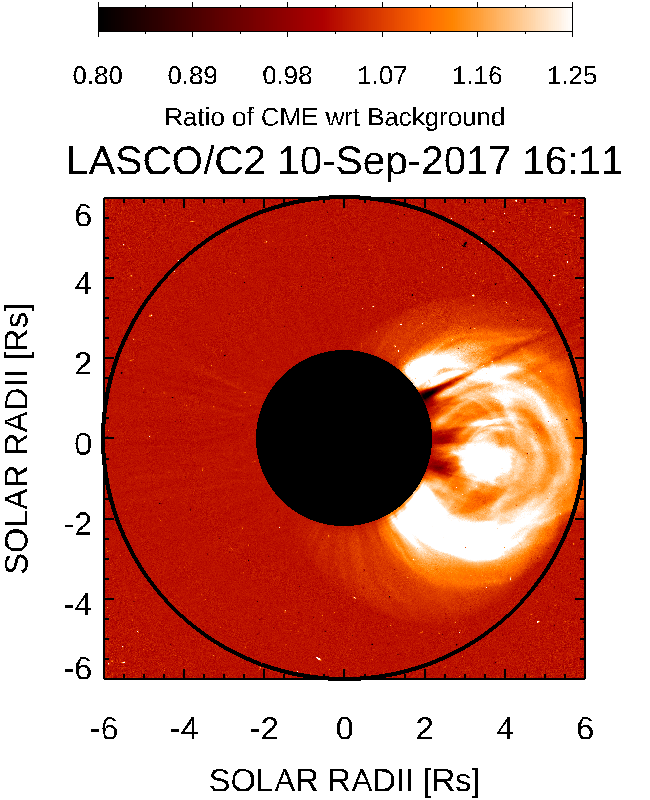}
\noindent\includegraphics[width=0.33\textwidth]{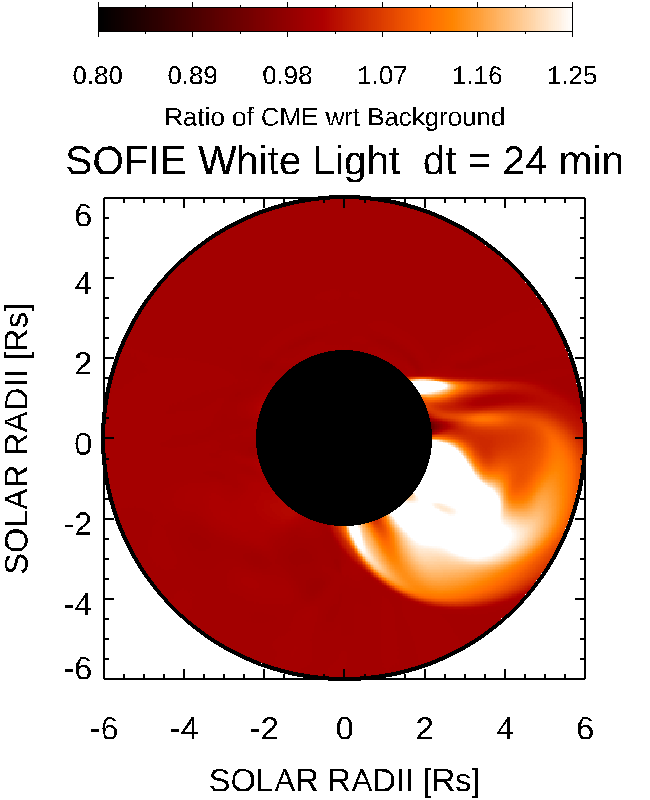}
\caption{In the same format as Figure~\ref{fig:cme_1} for the three events 2017-Sep-04, 2017-Sep-06, and 2017-Sep-10.}
\label{fig:cme_3}
\end{figure}

After obtaining the steady state background solar wind solution, we then launch the CME from the location of the parent active region by placing an imbalanced Gibson-Low \cite{Gibson1998} magnetic flux rope.
The parameters of the flux rope, including the total magnetic field, the flux rope size, and the flux rope orientation, are calculated based on the GONG magnetogram and the observed CME speed.
In Figures~\ref{fig:cme_1}, ~\ref{fig:cme_2}, and~\ref{fig:cme_3}, we show the 3D topology of the inserted flux rope (left column), the white-light image measured by the LASCO/C2 telescope (middle column), and the synthetic white-light image calculated from the simulation (right column).
In the left column, the surface of the Sun (at $1.105$ Rs) and a number of 3D magnetic field lines are colored according to the  radial component of the magnetic field. Note that the Sun and the magnetic field lines do not share the same color bar. The color bar shown in each plot represents the magnetic field strength on the magnetic field lines. The radial magnetic field on the Sun (at $1.105$ Rs) ranges from $-20$ Gauss to $20$ Gauss. 
The large scale magnetic field lines, besides the flux rope, are plotted to represent the overall structures of the coronal magnetic fields in each event. It is clearly seen that the field configurations differ dramatically from event to event. And the overall magnetic field strength on the solar surface also varies orders of magnitude. 
The perspective view of the Sun is that obtained from Earth.
Therefore, due to the projection effect, the flux rope of some events are not as distinguishable as the others, especially when the flux rope is located close to the center of the Sun, as viewed from Earth.

The middle and right columns of Figures~\ref{fig:cme_1}, ~\ref{fig:cme_2}, and~\ref{fig:cme_3} compare the white-light coronagraph observations (middle) and simulations (right) several tens of minutes after the eruption of each CME. The exact times shown in Figures~\ref{fig:cme_1}, ~\ref{fig:cme_2}, and~\ref{fig:cme_3} are selected on the basis of their clear CME detection in the LASCO/C2 field of view.
%that we chose to make the comparison are determined by the criterion when the CME is clearly seen from the field view of LASCO/C2 instrument. 
The exact time of the selected observational frame is shown in the title of each image. The images calculated from the simulation are chosen accordingly and the time, $dt$, after the CME eruption is shown. 

In the following, we briefly describe the white-light comparison of each individual CME between the observation and simulation.
In the 2012-Mar-07 event (top row of Figure~\ref{fig:cme_1}), the core structure of the CME compares well, and the leading edge of the CME reaches approximately the same radial distance between observation and simulation, although the overall expansion of the CME in the simulation is narrower than the observation, especially in the left flank. In the 2012-May-17 event (middle row of Figure~\ref{fig:cme_1}), the core structure, the leading edge, and the overall expansion of the CME are well-captured by the simulation. In the 2012-Jul-12 event (bottom row of Figure~\ref{fig:cme_1}), the CME is a halo CME (CDAW) and the flux rope originated from the center of the Sun as seen from Earth (left column). Therefore, the projection effect is large. From the LASCO/C2 image (middle column), the core structure of the CME has a southern part (the active region is located at S14W02) , which is captured in the simulation.

In the 2013-Apr-11 event (top row of Figure~\ref{fig:cme_2}), the core structure of the CME propagates toward the east as seen in the LASCO/C2 images. The envelope of the CME appears to be symmetric with respect to the solar equator. However, in the white-light image obtained from the simulation, the northern part of the CME is brighter than the southern part, demonstrating an extreme asymmetric shape. We examined the plasma properties in the low solar corona and found a high density region lying in front of the flux rope which slowed down the propagation of the CME and led to such an asymmetric structure.

In the 2014-Jan-07 event (middle row of Figure~\ref{fig:cme_2}), the CME erupted from the active region located at S15W11.
From the LASCO/C2 point of view, the CME was a halo CME but propagating mostly in the southwest direction. The initial simulation also obtains a halo (not shown here), which does not have the southwestern part as seen from the LASCO/C2 images. Therefore, it is very likely that the CME was deflected towards the west in the very early stage. We examined the magnetic fields around the active region where the flux rope was inserted and found there was a strong active region in the east of the flux rope. 
The CME eruption and propagation in this event has been analyzed in detail by \citeA{Mostl2015}.
They found the CME was “channeled” by strong nearby active region magnetic fields and open coronal fields into anon-radial propagation direction within $\sim$2.1 $R_s$.
In the current setup of simulations, since the initial speed of the CME was 2048 km s$^{-1}$, the flux rope is difficult to be deflected in the early stage. Therefore, in order to match that of the LASCO/C2 observation and also match the subsequent propagation of the CME, we shifted the location of the flux rope to the adjacent active region in the west, separated by 8$^{\circ}$ in longitude from the active region listed in Table~\ref{tbl:overview}. As seen from Figure~\ref{fig:cme_2}, the simulated CME propagates toward southwestern, which is comparable to the observations. However, the shifting of the flux rope to the west leads to issues when modeling the particle acceleration and propagation. 
%We will discuss it in more detail in Section~\ref{sec:mflampa}.

In the 2017-Jul-14 event (bottom row of Figure~\ref{fig:cme_2}), the white-light image from the observation and simulation is comparable, except that the CME shows a bright northern part in the simulation.
While in the observation, the core part of the CME leans toward the south.
The 2017-Sep-04 event (top row of Figure~\ref{fig:cme_3}) involved two CMEs. From the LASCO/C2 movie, there was a preceding CME eruption that occurred around 2 hours before the main CME, with a speed of 597 km s$^{-1}$ (CDAW). The previous CME propagated toward the west and the main CME took over the previous CME shortly after the eruption. In the LASCO/C2 image (top row of Figure~\ref{fig:cme_3}), we enclose the leading edge of the main CME for a better vision comparison with the simulation. 
In the simulation, we only launch the main CME. The radial distance of the CME leading edge and its propagation direction is in a good agreement with the observation.
Both the 2017-Sep-06 and 2017-Sep-10 events (middle and bottom rows of Figure~\ref{fig:cme_3}) show very good agreement between simulations and observations, in terms of the CME speed and propagation direction, including the interaction of the flux rope with the high density streamers in the background solar wind.

\subsection{Energetic Particles}\label{sec:mflampa}

Once the force-imbalanced flux rope was inserted into the active region, we run the coupled AWSoM-R and M-FLAMPA modules to solve the energetic particle acceleration and transport processes in the solar corona and inner heliosphere.
More than $600$ magnetic field lines are extracted from the 3D AWSoM-R solution.
The extracted magnetic field lines are followed in the local Lagrangian reference frame convecting with the solar wind plasma. A frequent (120 s) dynamic coupling between AWSoM-R and M-FLAMPA is performed to account for the propagation of the CME and CME-driven shock wave. 
In the simulation, the shock is identified by the sudden jump of the solar wind velocity along the extracted magnetic field lines. 
On each individual magnetic field line, the Parker diffusion equation is solved in the time-evolving Lagrangian coordinates. 
The diffusion strength close to the shock is determined by the total Aflv\'en wave intensity calculated self-consistently from the AWSoM-R simulation. The diffusion mean free path upstream of the shock, as described in \citeA{Sokolov2004}, is assumed to be a constant value, 0.3 AU. This setup is for simplicity and in the future simulations, the diffusion coefficients in the entire domain will be calculated from the AWSoM-R solution.
In this set of runs, perpendicular diffusion due to the field line random walk is not modeled.  
In modeling the nine SEP events, we followed $648$ magnetic field lines that cover 360$^o$ in longitude and -45$^o$ to 45$^o$ in latitude of the solar surface. The starting radial distance of the magnetic field lines is $2.5$ Rs, and the magnetic field lines are traced inward and outward until reaching the inner and outer boundaries. The starting points of the magnetic field lines are chosen to distribute uniformly in the sphere enclosed $2.5$ $R_s$. The latitudes of the active region that we insert the flux rope are within $\pm$ 17$^o$ around the solar equator. Therefore, a $\pm$ 45$^o$ coverage in latitudes is sufficient to calculate the particle flux in the ecliptic plane.

In this work, we are not trying the solve the injection problem, instead, we set the injection energy, $E_i$, in the shock system to be $10$ keV. The absolute level of the injected particles is determined by assuming a suprathermal tail ($\sim p^{-5}$) extending from the thermal momentum ($\sqrt{2mT}$) to the injected momentum ($p_i$) as follows \cite{Sokolov2004}:
  \begin{linenomath*}
  \begin{equation}
  f(p_i) = \frac{c_i}{2\pi}\frac{n}{(2mT)^{3/2}}\left(\frac{\sqrt{2mT}}{p_i}\right)^{5}
  \end{equation}
  \end{linenomath*}
where m is the proton mass, $n$ and $T$ are the local plasma density and temperature in energy units (if in Kelvins, $k_BT$ should stand instead, $k_B$ being the Boltzmann constant) calculated from AWSoM-R simulation. $c_i<1$ is the injection coefficient and $p_i$ is the injection momentum.
The physical meaning of the injection coefficient may be derived by integrating the assumed distribution of the suprathermal particles over momentum, which gives us their density: $4\pi\int\limits_{\sqrt{2mT}}^{p_i}fp^2\mathrm{d}p= c_in$. Hence, $c_i$ is a fraction of density of protons having suprathermal energy.
In order to compare with the observations, the injection level $c_i$ is adjusted for each individual SEP event.
These suprathermal particles are then accelerated on the magnetic field lines with negative velocity divergence ($\nabla\cdot\bf{u}<0$). The strength of the acceleration is fully dependent on the jump of plasma velocity, i.e. the shock strength \cite{Sokolov2004}.

\subsection{2D Distribution of Proton Flux}
\begin{figure}
\noindent\includegraphics[width=\textwidth]{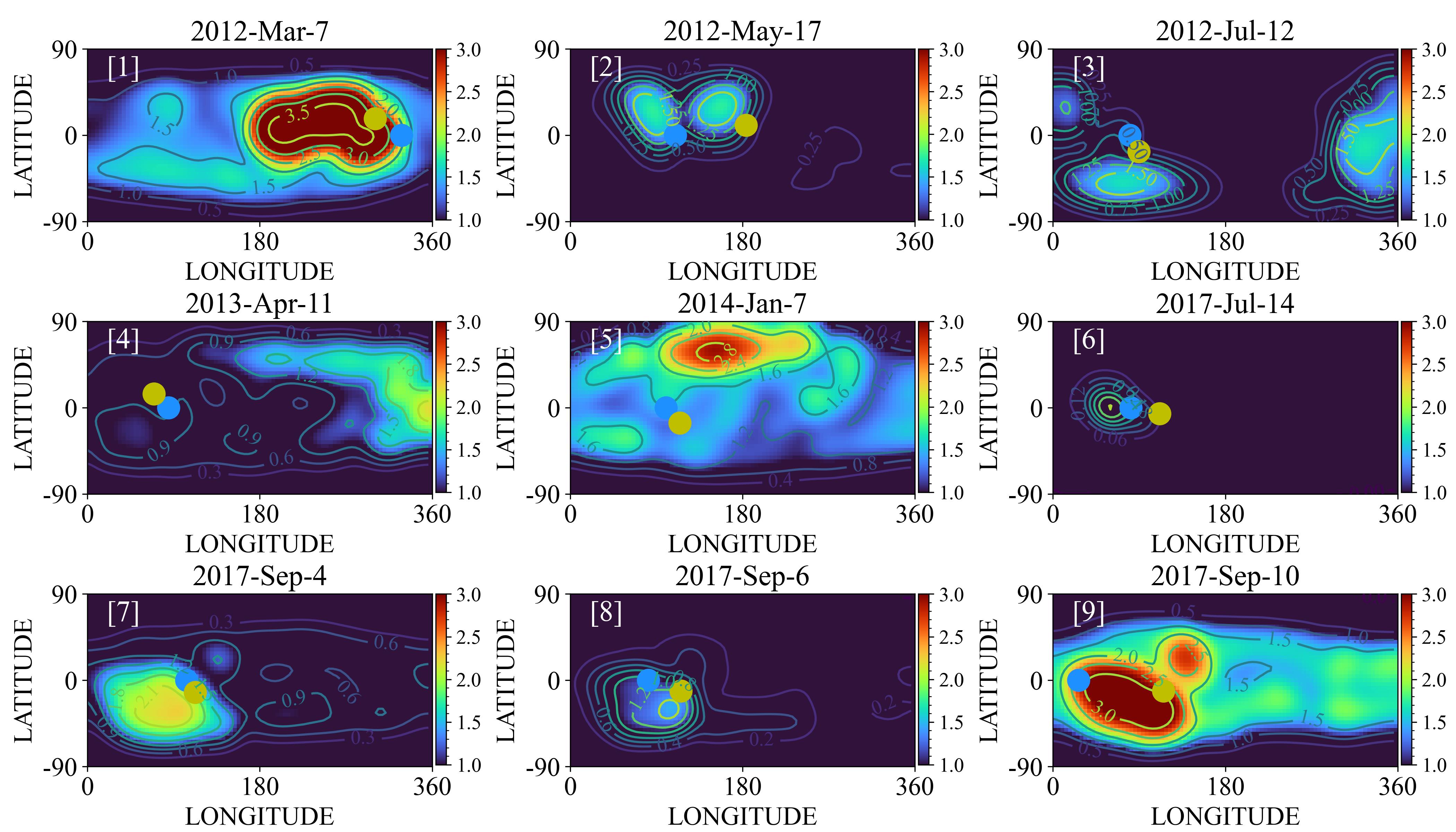}
\caption{2D distribution of energetic proton flux at energies greater than $10$ MeV. The proton flux is plotted in the logarithm scale. The nine events are plotted in the row-wise order.
The x and y axis shows the Carrington longitude and latitude on the sphere at 1 AU.
The locations of Earth are marked with blue solid circle, and the location of the inserted flux rope on the Sun are marked with yellow solid circle.}
\label{fig:sep_2d}
\end{figure}

Figure~\ref{fig:sep_2d} shows the 2D distribution of the logarithm of the energetic proton flux $1$ hour after the eruption of the CME flux rope, at energies greater than $10$ MeV. 
The x and y axis shows the Carrington longitude and latitude for a sphere at 1~AU.
Earth location is marked with a blue solid circle, and the location of the inserted flux rope on the Sun is marked with a yellow solid circle. The locations of the flux rope are marked in the plot to show the relative locations of Earth with respect to the CME, i.e. the source of energetic particles. Since the interplanetary magnetic fields follow Parker spiral in general \cite<e.g.>{Zhao2019}, the flux of energetic particles is distributed around 45$^o$ $\sim$ 65$^o$ eastern of the flux rope location, depending on the corona and interplanetary magnetic field configurations. In this set of runs, the injection coefficients are assumed to be uniform across the shock front (shock obliquity independent). Therefore, the 2D distribution of the energetic particles reflects the collective effect of the strength of the shock, the ambient plasma density and the temperature of the flux rope.

In the 2012-Mar-07 event, the parent CME erupted from the active region located at N17E15 (see Table~\ref{tbl:overview}), $15$ degree eastern of the Earth's longitude. The 2D proton flux distribution in Figure~\ref{fig:sep_2d} shows maxima around $90$ degree eastern of the Earth's location, which is consistent with the overall topology of the interplanetary magnetic fields. 
In the 2012-May-17 event, the parent CME erupted from the west limb, around $90$ degree western of the Earth's longitude. There are two local maxima in the 2D distribution of proton flux, which may be due to the non-uniform strength of the shock driven in front of the propagating flux rope that affects the acceleration process, or the variations of the ambient plasma properties that determines the suprathermal injection.

In the 2012-Jul-12 event, the parent CME erupted from near central meridian as seen from Earth. Since propagation direction of the CME leans toward the south, the proton flux in the southern hemisphere was also elevated due to the southern portion of the flux rope.
In the 2013-Apr-11 event, the parent CME erupted from active region located $12$ degree eastern of Earth, which is consistent with the 2D distribution of proton flux shown in Figure~\ref{fig:sep_2d}. As we discussed in Section~\ref{sec:cme}, the northern part of the CME is brighter than the southern part in the white-light image of the simulation, due to the high density region in front of the flux rope. Such an asymmetry structure was reflected in the 2D distribution plot of proton flux. The proton flux was elevated in the northern hemisphere and extended to a broader region than in the southern hemisphere, corresponding to a stronger particle source in the north.

In the 2014-Jan-07 event, the CME erupted from the active region located at S15W11. However, the 2D proton flux distribution shows local maxima far away from the expected region. This is due to the fine-tuning process that we performed in matching the white-light images between the observations and simulations as discussed in Section~\ref{sec:cme}. The flux rope was inserted to an active region to the west separated by 8 degrees in longitude from the active region that was responsible for the eruption. Meanwhile, the flux rope was also rotated in order to match the simulation with the observations, which leads to the unexpected northward propagation of flux rope. 
In the 2017-Jul-14 event, the parent CME erupted from S09W33, consistent with the 2D distribution of proton flux. Note that in panel [6] of Figure~\ref{fig:sep_2d}, Earth is very close to the center of the distribution.

The 2017-Sep-04, 2017-Sep-06, and 2017-Sep-10 are a sequence of events that their parent CMEs erupted from the same active region located at $16$, $34$, and $88$ degrees western of the Earth's longitude. As shown in the panels [7], [8], and [9] of Figure~\ref{fig:sep_2d}, the Earth's location was on the western, close to the center, and eastern of the energetic proton source. 

The 2D distribution of the energetic proton flux highly depends on the shock properties, i.e. shock strength, along the connected magnetic field lines with the corresponding CME. 
Furthermore, the absolute particle flux is determined by the number of seed particles that are injected into the shock system.
In plotting the 2D distributions shown in Figure~\ref{fig:sep_2d}, we varied the injection coefficient for each individual event in order to obtain comparable results with the observations made by GOES satellite. 
The relative injection ratio is summarized in Table~\ref{tbl:injection} and will be discussed in detail below.
Note that for some events, the injection coefficient is much larger than $1$, e.g. the 2012-Mar-07 event and 2014-Jan-07 event. There are many reasons that could lead to such large injection coefficients. One of the reasons is the underestimation of the pre-existing seed particle sources at the event eruption, including the preceding CMEs and the flares. Another factor that will affect the injection coefficient is the combined effect of the magnetic connectivity between the CME shock front and the earth's location with neglecting the perpendicular diffusion in the calculation. A small displacement of the earth's magnetic footpoint with respect to the shock front, together with an overestimation/underestimation of the CME shock properties will lead to a large variation of the proton flux. In this work, the perpendicular diffusion is not modeled, therefore, the proton flux contribution from cross-field diffusion, which is very important for poorly-connected events, is missing.

 \begin{table}
 \caption{Injection Coefficients of the nine SEP events}
 \label{tbl:injection}
 \centering
 \begin{tabular}{cc}
 \hline
  Event  & Injection Coefficient ($c_i$)  \\
 \hline
   2012-Mar-07 & 5 \\
   2012-May-17 & 0.025 \\
   2012-Jul-12 & 0.025 \\
   2013-Apr-11 & 1.25 \\
   2014-Jan-07 & 2.5 \\
   2017-Jul-14 & 0.00025 \\
   2017-Sep-04 & 0.25 \\
   2017-Sep-06 & 0.025 \\
   2017-Sep-10 & 1.25 \\
 \hline
 \end{tabular}
 \end{table}

\subsection{Time Profiles}
\begin{figure}
\noindent\includegraphics[width=\textwidth]{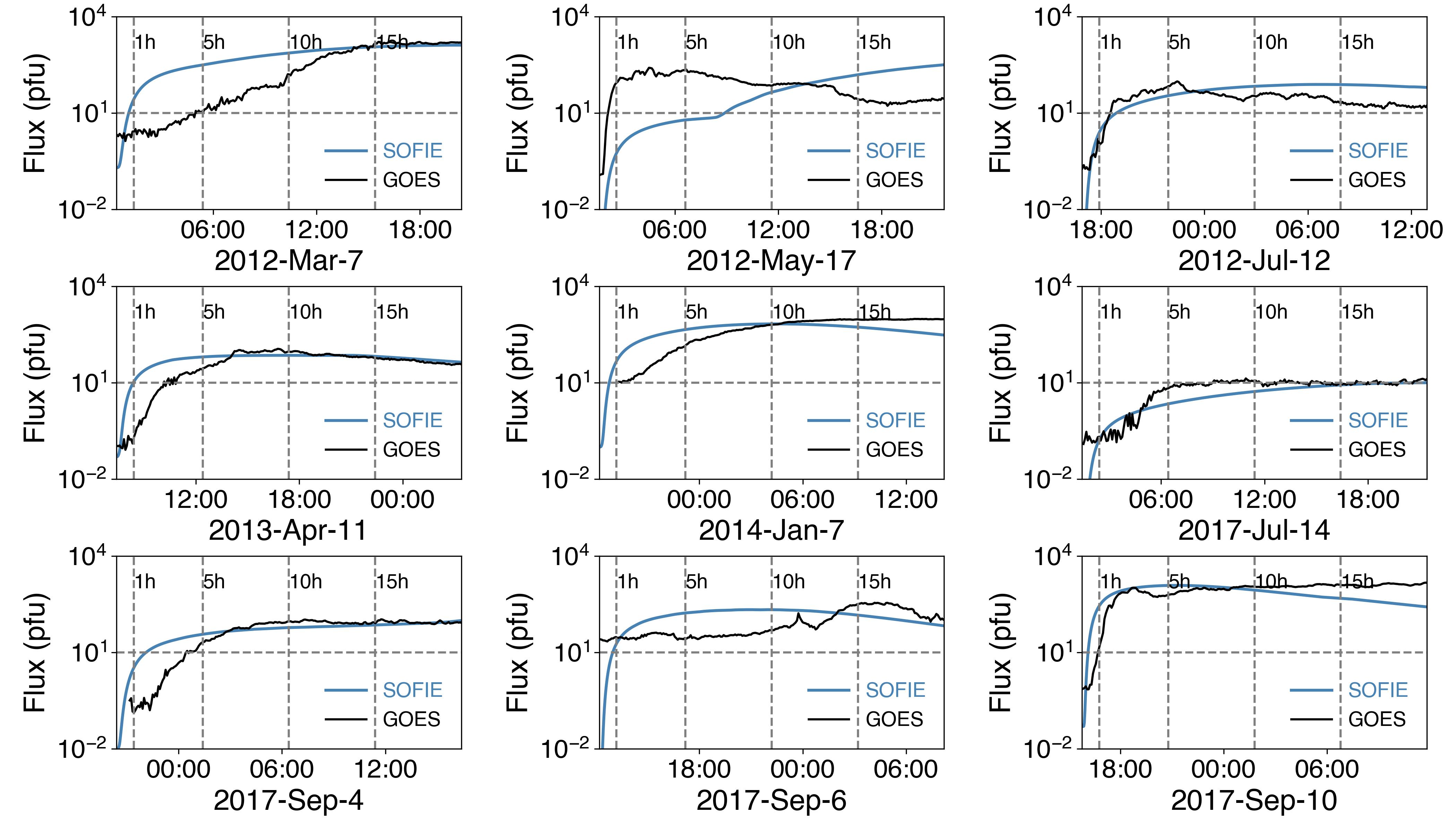}
\caption{The comparison of proton flux at energies greater than 10 MeV between observations (black) and simulation (blue). Nine events are plotted in the row-wise order. The horizontal dashed line represents the threshold of $10$ pfu and the four vertical dashed lines represent $1$h, $5$h, $10$h, and $15$h after the CME eruption. A total time period of 20 hours after the CME eruption is shown.}
\label{fig:flux}
\end{figure}

Figure~\ref{fig:flux} compares proton intensities measured by  GOES with the time dependent flux profiles obtained from the simulations. 
The flux profiles are calculated by extracting the $>10$ MeV proton flux at Earth's location from series of 2D particle distributions as shown in Figure~\ref{fig:sep_2d}.
A total of $20$ hours are plotted. The horizontal dashed lines represent the $10$ particle flux unit (pfu) threshold used by agencies to determine whether the radiation caused by the energetic protons raises any concern. The four vertical dashed lines indicate the times $1$h, $5$h, $10$h, and $15$h after the eruption of the CME flux rope.
As we mentioned above, the absolute proton flux is multiplied by a factor of the injection coefficient in order to get comparable match between  observations and simulations. Therefore, in the following discussion, we focus on the rising phase and relative level of the flux profiles.

Based on the relative location of Earth with respect to the source of energetic protons, a prompt onset of protons is expected for the events when Earth is well-connected to the source of energetic protons. While the proton flux is expected to show a gradual increase if Earth's location falls outside of the particle source. As shown in the 2D distribution of energetic protons (Figure~\ref{fig:sep_2d}), in most of these events, Earth's location is on the edge of the particle distribution at 1~AU, including the 2012-Mar-07, 2012-May-17, 2017-Jul-14, 2017-Sep-04, 2017-Sep-06, and 2017-Sep-10 events. In the 2012-Jul-12, 2013-Apr-11, and 2014-Jan-07 events, Earth location is far away from the particle distribution at 1~AU. The change of the proton flux with time, especially in the early phase, depends on the time evolution of the CME flux rope, together with the change of magnetic connectivity between Earth and the CME.

The comparison between the simulations and observations shown in Figure~\ref{fig:flux} displays some discrepancies. 
A number of factors could contribute to these discrepancies. One of them  is the background solar wind medium where the CME flux rope and energetic protons propagate. The solar wind background in this work is a steady-state solution driven by the solar magnetic fields measured at a single time (before the flare eruption) and the 3D solar wind solution has been compared to measurements obtained from a single near-Earth point in space that might not be representative of all the medium sampled by the particles as they propagate from the CME shock front to Earth. And the solar wind disturbances, including ICMEs, which are abundant during solar maximum, are not modeled. A second factor is due to the fact that the longitudinal extent of the shock may be underestimated/overestimated. Our CME flux-rope  white-light simulation images have been validated with plane-of-sky images of the LASCO/C2 observation that do not include the extent of the CME in longitude. A third factor is the assumption of the same constant parallel mean free path in all SEP events and the lack of cross-field diffusion processes when modeling energetic particle transport in interplanetary space. Keeping these factors in mind, we discuss the comparisons between simulations and observations for all the events in details below.

In the 2012-Mar-07 event, the proton flux calculated from the simulation shows a prompt increase, which is different from the gradual increase in the observation. This may due to the CME-driven is narrower in the observation than in the simulation. 
The injection coefficient is estimated to be 5. 
As discussed in Section~\ref{sec:overview}, there are two CME eruptions associated with this event, and the energetic particles from these two eruptions merged together after the two clear onset phases. Therefore, the injection coefficient, $5$ for this event, may reflect the contribution of the two eruptions.
Besides, the $>10$ MeV proton flux was already elevated before the onset of this event from the observations. The pre-event elevated proton flux is due to a CME eruption that occurred on 2012 Mar 4 at 11:00:07 UT (CDAW). 

In the 2012-May-17 event, the onset phase time matches well between the observation and simulation. The second enhancement of proton flux at around $7$ hours after the CME eruption was due to the CME evolution and the fact that Earth's magnetic connectivity changed establishing connection with a region with larger proton flux. Due to the second enhancement of the proton flux, the injection coefficient for this event does not reflect the difference of the overall level of proton flux between simulation and observation. 

In the 2012-Jul-12 event, the timing of proton flux in the simulation matches very well with the observations, especially in the early phase. The mismatch of the declining of the proton flux after $10$ hours may due to the assumption of the mean free path in the simulation. The effect of the mean free path on the decay phase of the proton flux will be discussed below.

In the 2013-Apr-11 event, the calculated proton flux shows a quicker onset phase than the observations. The slower onset may due to the poor magnetic connection of Earth to the CME (with an AR of N09E12). The proton flux after $6$ hours between observation and simulation matches quite well and the injection coefficient of $1.25$ is a reasonable value.

The 2014-Jan-07 is a special case, as we discussed above. The 2D proton flux distribution shows the particle source is far away from the expected region, due to the fine-tuning processes of the inserted flux rope. Moreover, the $>10$ MeV proton flux in the observation was well-above the background due to a previous eruption that occurred at 08:00~UT on 2014 January 06.

The gradual onset phase in the 2017-Jul-14 event matches well between observation and simulation. 
The injection coefficient in this event is estimated to be $2.5\cdot10^{-4}$.
This small value of injection could be due to the slower speed of the parent CME, $750$ km s$^{-1}$. 
%There are less suprathermal protons that are injected into the shock acceleration. 
However, the CME speed in the 2013-Apr-11 event is $743$ km s$^{-1}$, comparable to the one in the 2017-Jul-14 event, but the 2013-Apr-11 event has an injection coefficient of $1.25$.
Another reason for the small injection coefficient is that the eruption of the 2017-Jul-14 event was near solar minimum, when the solar activity was low, and the remnant population of prior SEP events that could act as seed particle population for the processes of particle acceleration at the shock could also be low.
%The overall magnetic field strength on the solar surface is relatively large in this event (last row in Figure~\ref{fig:cme_2}), resulting in a more efficient acceleration in the corona shocks.

The 2017-Sep-04, 2017-Sep-06, and 2017-Sep-10 are a series of events that their parent CMEs erupted from the same active region. The injection coefficients in these three events are $0.25$, $0.025$, and $1.25$. 
The CMEs associated with the 2017-Sep-04 event are twin-CMEs \cite{Li2012} as we discussed in Section~\ref{sec:overview} and shown in Figure~\ref{fig:cme_3}. The more efficient acceleration in the twin-CME system \cite{Li2012,Zhao2014,Ding2013} could be one of the potential reasons why the injection coefficient in this event is much larger than the 2017-Jul-14 event, although this event occurred under solar minimum conditions.
The 2017-Sep-06 event occurred in the decay phase of the 2017-Sep-04 event. Therefore, the onset phase between the observation and simulation does not compare well. 
The onset phase in the 2017-Sep-10 event calculated from the simulation is faster than the observation. This may due to the overall extension of the CME flux rope and the magnetic connectivity at the beginning of the event. Similar to the 2012-Jul-12 event, the declining phase in the simulation is faster than the simulation, indicating a faster deceleration of the CME in the simulations or a larger mean free path assumption. 

%Except for the 2012-Mar-07 and 2017-Jul-14 events, the injection coefficient determined in all other events are within one order of magnitude of the default value.
The determination of the injection coefficient in each individual event is affected by the properties of the shocks driven by the eruption of the CME flux rope, including the spatial extension of the shock surfaces and the strengths of the shocks.
Hence, the value of the injection coefficient does not necessarily imply there are more or less suprathermal protons, in the energy of $10$ keV, that are accelerated in the diffusive shock acceleration process. 
An estimation of a larger CME flux rope or a stronger CME-driven shock will lead to a smaller injection coefficient and vice versa.
Besides, the magnetic connectivity between the Earth's location and the CME shock front also affect the injection coefficient. 
If the Earth's location is close to the edge of the particle source, a small change of the size of the CME flux rope or a little error in the magnetic connectivity calculation will result in a larger or smaller injection coefficient. 
From Figures~\ref{fig:cme_1}, \ref{fig:cme_2}, and \ref{fig:cme_3}, the comparison between the simulation and observation is only performed for the SOHO observations, which include a large projection effect. In the future work, a multi-spacecraft validation of the white-light CME image will be included. 
Moreover, together with C2 observation, C3 observation will also be used to monitor the acceleration or deceleration of the CME flux rope in the solar corona. 
This is because the onset phase contains competing processes between the continuous acceleration of protons and the diffusion process. A significant deceleration of the CME flux rope propagation in the very early phase would reduce the acceleration efficiency of energetic protons, especially in the larger energy end.

\subsection{Decay Phase}
\begin{figure}
\noindent\includegraphics[width=0.5\textwidth]{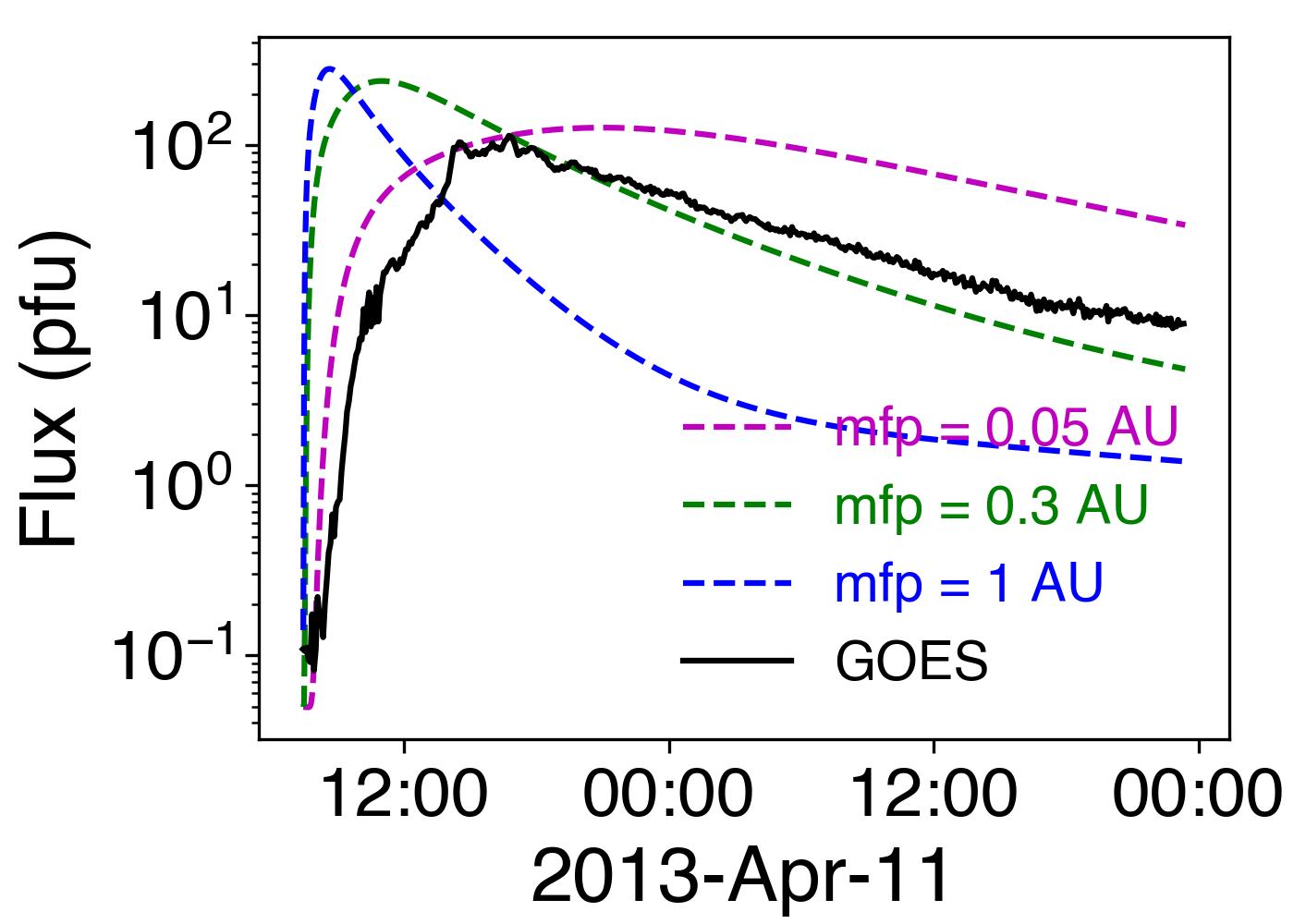}
\caption{The effect of far-upstream mean free paths on the calculated proton flux profiles in the 2013-Apr-11 event. The GOES observation is plotted in black. The calculated proton flux profiles with different mean paths (mfp) are plotted in magenta (mfp=0.05 AU), green (mfp=0.3 AU), and blue (mfp=1 AU). }
\label{fig:mfp}
\end{figure}

The ambient solar wind plasma properties affect the transport of energetic particles, including the magnetic field turbulence. The timing of the first arriving particles, the timing when the particle crosses the preset threshold, \cite{Wang2015,Qin2006} e.g. $10$ pfu, and the time dependent and event-integrated energy spectra \cite{Zhao2016,Zhao2017} are all impacted by the magnetic field turbulence. In the simulation, the ambient solar wind plasma is calculated by running the steady-state MHD simulation using Stream-Aligned AWSoM-R module in SWMF. The mean free path upstream of the shock is assumed to be $0.3$ AU in all of the nine simulations, for simplicity. In Figure~\ref{fig:mfp}, we show the effect of different mean free paths on the proton flux profiles for the 2013-Apr-11 event as an example. The magenta, green, and blue dashed curves show the flux profiles with far-upstream mean free paths of $0.05$ AU, $0.3$ AU, and $1$ AU. The calculated proton fluxes are extracted from a sample magnetic field line. Both the onset phase and the decay phases depend on the value of mean free paths in the three cases as expected. Employing the turbulence strength calculated from the MHD simulation is one of the future steps to improve the SOFIE model.

\section{Discussion}
In this paper, we describe the physics-based SEP model, SOFIE, and its application in modeling nine historical SEP events.
The simulations of the SEP events start from calculating the background solar wind using the AWSoM-R model, in which the solution of the solar wind plasma is driven by the measurement of the Sun's magnetic field. 
The acceleration of energetic protons in SOFIE is solved in the CME-driven shock generated by the eruption of CME flux rope. 
The CME is modeled by inserting an imbalanced flux rope on the corresponding active region on the Sun using the EEGGL model.
The acceleration and transport of energetic protons are modeled using the M-FLAMPA model, in which the Parker diffusion equations are solve along individual time-evolving magnetic field lines. In such regards, SOFIE model is a data-driven and self-consistent SEP model. 

In this work, we perform a systematic test of using SOFIE model to simulate SEP events. The steady-state background solar wind macroscopic properties (radial solar wind speed, number density, temperature, total magnetic field strength) calculated from the AWSoM-R is compared and validated against in-situ measurements. The white-light coronagraph image of the erupted flux rope generated by the CME generator, EEGGL, is compared and evaluated with SOHO/LASCO/C2 observations. This is only a single-observer comparison, therefore, the longitudinal extent of the flux rope has not been compared to observations. The proton flux at energies greater than $10$ MeV calculated by M-FLAMPA is compared with GOES observation for the first $20$ hours. In order to obtain a comparable flux level with observations, different injection coefficients are used for each event. The potential factors that may affect the injection coefficient include the multiple CME eruptions in one SEP event, the elevated suprathermal particles from previous eruptions, and solar activity level. We also discussed the effect of the upstream mean free path on proton flux profiles, especially the declining phase. In the current set of runs, the upstream mean free paths are assumed to be the same for all the events for simplicity. This assumption may lead to a faster or slower declining profile in the simulation. The mean free paths may also affect the onset phase of the SEP event, making it more difficult to evaluate the acceleration/deceleration of CME propagation in the early stage.

The most time and resources consuming part of the SOFIE model is when modeling the propagation of the CME flux rope in the solar corona domain (1.05 $R_s$ to 20 $R_s$). In this stage, the SOFIE model runs at the same speed as real-time with $2000$ cpu cores. It can run faster than real-time if more cpu cores are used. When the CME flux rope leaves the solar corona domain, several hours after the CME eruption, SOFIE model runs much faster than real-time, thus empowering the capability of using SOFIE model in predicting the properties of SEP events.

The necessity of transporting energetic particles in the solar wind solution calculated from an MHD simulation is due to the complex physical processes therein.
The transport of energetic particles in interplanetary space involves many different physical processes, including adiabatic cooling, magnetic focusing, as well as parallel and perpendicular diffusion. All these processes depend on the properties of ambient solar wind background.
The magnetic field turbulence affects the timing of the first arriving particles, the timing when the particle flux crosses a pre-set threshold \cite{Wang2015, Qin2006}, and the time-dependent and event-integrated energy spectral index \cite{Zhao2016, Zhao2017}. In the set of runs in this work, the upstream mean free paths are assumed to be constant and the effect of magnetic turbulence strength from the AWSoM-R simulation will be discussed in subsequent papers.

Besides the steady-state background solar wind, CMEs and ICMEs, which are the main accelerators of energetic particles travel through the ambient solar wind medium, 
interact  with its surrounding plasma and magnetic field, causing significant distortions and disruptions of the solar wind plasma \cite{Manchester2004, Manchester2004a, Manchester2005, Manchester2008, Manchester2012}. These distortions affect the acceleration and transport of energetic particles. There are also SEP events that are associated with more than one CME eruption, e.g the 2012-Mar-07 and 2017-Sep-04 events. The underlying acceleration of energetic particles is likely to be enhanced according to the twin-CME scenario \cite{Li2012,Zhao2014,Ding2013}. 
In this work, when modeling the nine historical SEP events, each event is only associated with one CME eruption and the simulation of the background medium does not include prior CMEs that could affect the transport of SEPs. In future work, we will examine the performance of SOFIE in modeling more than one CME eruption.

\section*{Open Research Section}
The in-situ solar wind plasma properties used in this work is available in the Space Physics Data Facility \url{https://spdf.gsfc.nasa.gov/}. The white-light image data is available in the SOHO/LASCO website \url{https://lasco-www.nrl.navy.mil/index.php?p=content/retrieve/products}. The GOES data is available at \url{https://www.ngdc.noaa.gov/stp/satellite/goes/index.html}. All the simulation data including the 3D steady-state solution of the solar wind plasma, the 2D white-light image data, the 2D distribution of protons, and the time dependent flux profiles are publicly available at the Deep Blue Data Repository maintained by the University of Michigan \url{https://deepblue.lib.umich.edu/data/concern/data_sets/cn69m504s}.

\acknowledgments
This work was supported in part by NASA LWS Strategic Capabilities (SCEPTER) project at the University of Michigan under NASA grant 80NSSC22K0892, NASA SWxC grant 80NSSC23M0191 (CLEAR), NASA LWS grant 80NSSC21K0417, NASA R2O2R grant 80NSSC22K0269, NASA HSR grant 80NSSC23K0091, NSF ANSWERS grant GEO-2149771.
D.L. acknowledges support also 
from NASA Living With a Star (LWS) program
NNH19ZDA001N-LWS, and the Heliophysics
Innovation Fund (HIF) program of the Goddard Space Flight Center.
We thank the ACE SWEPAM instrument team and the ACE Science Center for providing the ACE data.
We thank the SOHO project, an international cooperation between ESA and NASA.
We thank the GOES team.
This work utilizes data from the National Solar Observatory Integrated Synoptic Program, which is operated by the Association of Universities for Research in Astronomy, under a cooperative agreement with the National Science Foundation and with additional financial support from the National Oceanic and Atmospheric Administration, the National Aeronautics and Space Administration, and the United States Air Force. The GONG network of instruments is hosted by the Big Bear Solar Observatory, High Altitude Observatory, Learmonth Solar Observatory, Udaipur Solar Observatory, Instituto de Astrofísica de Canarias, and Cerro Tololo Interamerican Observatory.
Resources supporting this work were provided, in part, by the NASA High-End Computing (HEC) Program through the NASA Advanced Supercomputing (NAS) Division at Ames Research Center.
The authors acknowledge the Texas Advanced Computing Center (TACC) at The University of Texas at Austin for providing HPC resources that have contributed to the research results reported within this paper. \url{http://www.tacc.utexas.edu}

\end{document}